\documentclass[journal]{IEEEtran}
\usepackage{latexsym,amsfonts,amssymb,amsthm, setspace, verbatim,cite,mathrsfs,graphicx,bm,stfloats, hyperref, tikz,xcolor,soul}
\usepackage[cmex10]{amsmath}
\usetikzlibrary{arrows,shapes,backgrounds,fit}
\tikzset{leftvertex/.style={rectangle, fill=purple, inner sep=0pt, minimum size=2mm},
rightvertex/.style={rectangle, fill=orange, inner sep=0pt, minimum size=2mm}}
\tikzset{Sleftvertex/.style={rectangle, fill=purple, inner sep=0pt, minimum size=1mm},
Srightvertex/.style={rectangle, fill=orange, inner sep=0pt, minimum size=1mm}}
\pgfdeclarelayer{background}
\pgfdeclarelayer{foreground}
\pgfsetlayers{background,main,foreground}

\newcommand{\be}{\begin{equation}}
\newcommand{\ee}{\end{equation}}
\newcommand{\ben}{\begin{equation*}}
\newcommand{\een}{\end{equation*}}

\newtheorem{defi}{Definition}
\newtheorem{thm}{Theorem}
\newtheorem{fact}{Fact}

\newcounter{mytempeqncnt}
\begin{document}
\title{A New Achievable Rate Region for the Multiple-Access Channel
  with Noiseless Feedback}
\author{Ramji Venkataramanan,~\IEEEmembership{Member,~IEEE,}
S.~Sandeep Pradhan,~\IEEEmembership{Member,~IEEE,}

\thanks{Manuscript received; revised. This work was supported by NSF grants CCF-0448115 (CAREER),  CCF-0915619. The material in this paper was
presented in part at the IEEE International Symposium on Information Theory, Seoul, South Korea, June 2009.}%
\thanks{R.~Venkataramanan was with the Department of Electrical Engineering and Computer Science, University of Michigan.
He is now with the Department of Electrical Engineering, Yale University, New Haven, CT 06511, USA (e-mail:rvenkata@umich.edu).}%
\thanks{S. Sandeep Pradhan is with the Department of Electrical Engineering and
Computer Science, University of Michigan, Ann Arbor, MI 48109, USA (e-mail:pradhanv@eecs.umich.edu).}
\thanks{Communicated by M.~Gastpar, Associate Editor for Shannon Theory.}
}
\maketitle
\begin{abstract}
A new single-letter achievable rate region is proposed for the two-user discrete
memoryless multiple-access channel(MAC) with noiseless feedback. The proposed region includes the Cover-Leung rate region \cite{CL81}, and it is shown that the
inclusion is strict. The proof uses a block-Markov superposition strategy based on the observation that the messages of the two users
are correlated given the feedback. The rates of transmission are too high for each encoder to decode the other's message directly using the
feedback, so they   transmit correlated information in the next block to  learn  the message of one another.  They  then cooperate in the following block to resolve the residual uncertainty of the decoder. The coding scheme may be viewed as a natural generalization of the Cover-Leung scheme
with a delay of one extra block and a pair of additional auxiliary random variables. We compute the proposed rate region for two different MACs and
compare the results with other known rate regions for the MAC with feedback. Finally, we show  how the coding scheme can be extended to obtain
larger rate regions with more auxiliary random variables.
\end{abstract}

\section{Introduction}
\label{sec:intro}
\begin{figure}
\centering
\includegraphics[width=3.5in]{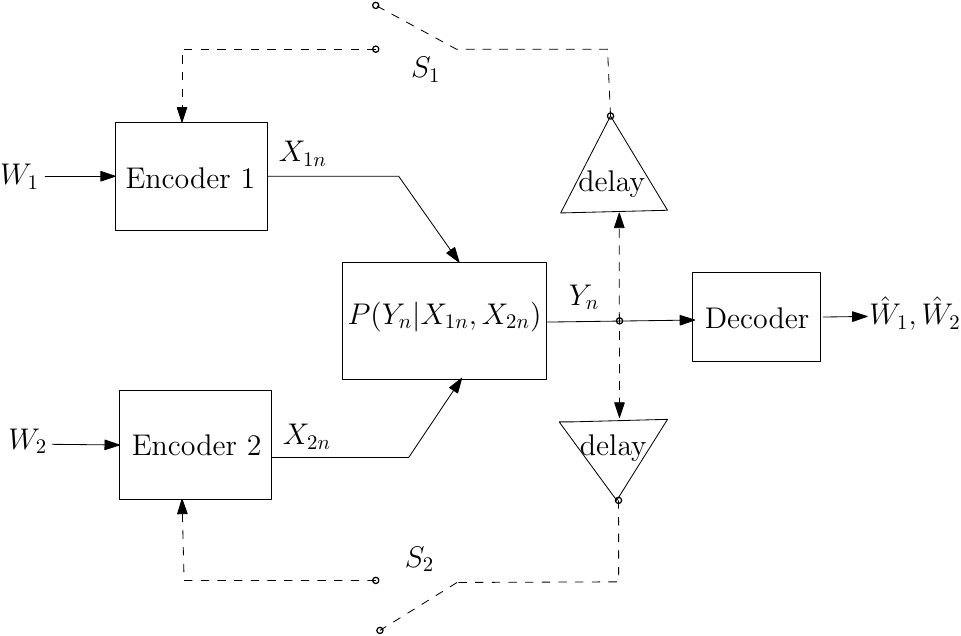}
\caption{The multiple-access channel. When $S_1,S_2$ are closed there is feedback to both encoders.}
\label{fig:macfb}
\end{figure}

\IEEEPARstart{T}he two-user discrete memoryless multiple-access channel (MAC) is shown in Figure \ref{fig:macfb}. The channel has two inputs $X_1,X_2$, one output $Y$, and is characterized by a conditional probability law $P_{Y|X_1X_2}$. A pair of transmitters wish to reliably communicate independent information to a receiver by using the channel simultaneously. The transmitters each have access to one channel input, and the receiver has access to the channel output. The transmitters do not communicate with each other. The capacity region for this channel without feedback ($S_1$ and $S_2$ open in Figure \ref{fig:macfb}) was determined by Ahlswede \cite{Ahls71} and Liao \cite{Liao72}.

In a MAC with noiseless feedback, the encoders have access to all previous channel outputs before transmitting
the present channel input. Gaarder and Wolf \cite{GaardWolf75} demonstrated that feedback can
enlarge the MAC capacity region  using the example of a binary erasure MAC.  Cover and Leung \cite{CL81} then established
a single-letter achievable rate region for discrete memoryless MACs with feedback. The Cover-Leung (C-L)
region was shown to be the feedback capacity region for a class of discrete memoryless MACs \cite{Willems82}. However, the C-L region is
smaller than the feedback capacity in general, the white Gaussian MAC being a notable example \cite{Ozarow84,Wigger08}. The feedback capacity region of
the additive white Gaussian MAC was determined in \cite{Ozarow84} using a Gaussian-specific scheme; this scheme is an extension of
the  Schalkwijk-Kailath scheme \cite{SchalkwijkKailath66} for the point-to-point white Gaussian channel with feedback.  The capacity region
of the MAC with feedback was characterized by Kramer \cite{Kramer98,Kramer03} in terms of directed information. However, this is a `multi-letter' characterization and is not computable. The existence of a single-letter capacity characterization for the discrete memoryless MAC with feedback remains an open question. A single-letter extension of the C-L region was proposed by Bross and Lapidoth in \cite{BrossLapi05}. Outer bounds to the capacity region of the MAC with noiseless feedback were established in \cite{hekstra89} and \cite{tandon09}. In \cite{An09}, it was shown that the optimal transmission scheme
for the MAC with noiseless feedback could be realized as a state machine, with the state at any time being the
a posteriori probability distribution of the messages of the two transmitters.

 MACs with partial/noisy feedback have also been considered in several papers. Willems \cite{willems84} showed that the C-L rate region can be achieved even with partial feedback, i.e., feedback to just one decoder. Achievable regions for memoryless MACs with noisy feedback were
obtained by Carleial \cite{carleial82} and Willems \cite{willems83}; outer bounds for this setting were obtained in \cite{gastKra06}. Recently, improved achievable rates for the Gaussian MAC with  partial or noisy feedback were derived in \cite{LapiWigger09}.

The  basic idea behind reliable communication over a MAC with feedback is the following. Before communication begins, the two transmitters
have independent messages to transmit. Suppose the transmitters use the channel once by sending a pair of channel inputs which are functions of the
corresponding messages. Then, conditioned on the channel output, the messages of the two transmitters become statistically correlated. Since
the channel output is available  at all terminals before the second transmission, the problem now becomes one of transmitting correlated messages over the MAC. As more channel uses are expended,  the posterior correlation between the messages increases. This correlation can be exploited to combat interference and channel noise more effectively in subsequent channel uses.  The objective is to capture this idea quantitatively  using a single-letter information-theoretic characterization.

The Gaarder-Wolf and the C-L schemes exploit feedback in two stages. Each message pair is conveyed to the decoder over two successive blocks of transmission. In the first block, the two  encoders transmit messages at rates outside the no-feedback capacity region.  At the end of this block, the decoder cannot decode the message pair; however, the rates are low enough for each encoder to decode the message of the other using the feedback.  This is possible because each encoder has more information than the decoder. The decoder now forms a list of highly likely pairs of messages.  The two encoders can then cooperate and send a common message to resolve the decoder's list in the next block. In the C-L scheme, this procedure is repeated over several blocks, with fresh information
superimposed over resolution information in every block. This block-Markov superposition scheme yields a single-letter achievable rate region for the MAC with feedback. In this scheme, there are two kinds of communication that take place: (i) Fresh independent information  exchanged  between the encoders, (ii) Common resolution information communicated to the receiver. This scheme provides a strict improvement
over the no-feedback capacity region.

Bross and Lapidoth \cite{BrossLapi05} obtained a single-letter inner bound to the capacity  rate region by constructing a novel coding scheme which uses the C-L scheme as the starting point. In their scheme, the two encoders spend additional time at the end of each block to engage in a two-way exchange, after which they are able to perfectly reconstruct the messages of one another. In the next block, the encoders cooperate to send the common resolution
information to the decoder. This coding scheme reduces to the C-L scheme when there is no two-way exchange.

In this paper, we propose a new achievable rate region for the MAC with feedback
 by taking a different path, while still using C-L region
 as the  starting point. To get some insight into the proposed approach,
 consider a pair of transmission rates significantly larger than any rate pair in the
 no-feedback capacity region, i.e., the rate pair is
outside even the C-L rate region. Below we describe a three-phase scheme to communicate at these rates.

 \emph{First Phase}: The encoders transmit independent information at the chosen rates over the channel in the first phase, and receive the corresponding block of channel  outputs via feedback. The rates are too high for each encoder to correctly decode the message of the other. At the end of this phase, encoder $1$ has its own message, and a list  of highly likely  messages of encoder $2$. This list is  created by collecting all the  $X_2$ sequences that are compatible (jointly typical) with  its own channel input and the channel output,  i.e., the $(X_1,Y)$ sequence pair. 
In other words, the list is a high conditional probability subset of the set of  messages of encoder $2$; this set is clearly smaller than the original message set of encoder $2$. Similarly, encoder $2$ can form  a list of highly likely  messages of  encoder $1$. Thus at the end of the first phase, the encoders have correlated information. They wish to transmit this information over the next block.

\begin{figure}
\centering
\includegraphics[width=3.4in]{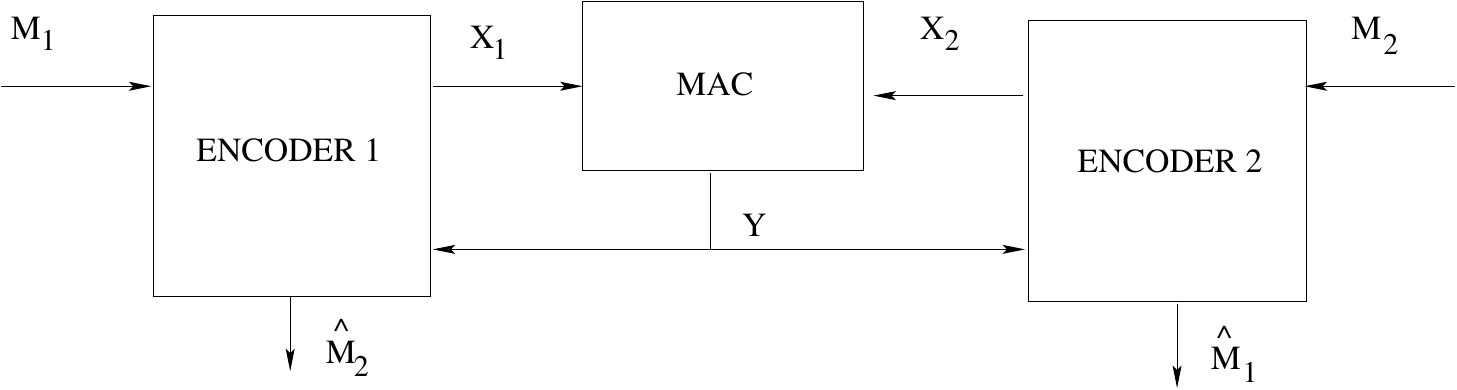}
\caption{First phase: transmission of independent information on
  common output two-way channel with list decoding}
\label{fig:phase_one}
\end{figure}

Conditioned on the channel output sequence, the above lists of the two encoders together can be thought of as a high-probability
subset of $\mathcal{M}_1 \times \mathcal{M}_2$, where $\mathcal{M}_1$ and $\mathcal{M}_2$ denote the message sets of the two encoders.
A useful way to visualize this is in terms of a bipartite graph: the left vertices of the graph are the
encoder $1$ messages that are compatible with the $Y$ sequence, and the right vertices are the
encoder $2$ messages that are compatible with the $Y$ sequence. A left vertex and a right vertex are connected by an edge if
and only if the corresponding messages are \emph{together} compatible with the $Y$ sequence, i.e.,
the corresponding $(X_1,X_2)$ sequence pair is jointly typical with the  $Y$ sequence.  This  bipartite graph
(henceforth called a message graph) captures the decoder's uncertainty about the messages of the two encoders.
In summary, the first phase of communication can be thought of as transmission of independent information by two
terminals over a common-output two-way channel with list decoding, as shown in Figure \ref{fig:phase_one}.

\begin{figure}
\centering
\includegraphics[width=3.4in]{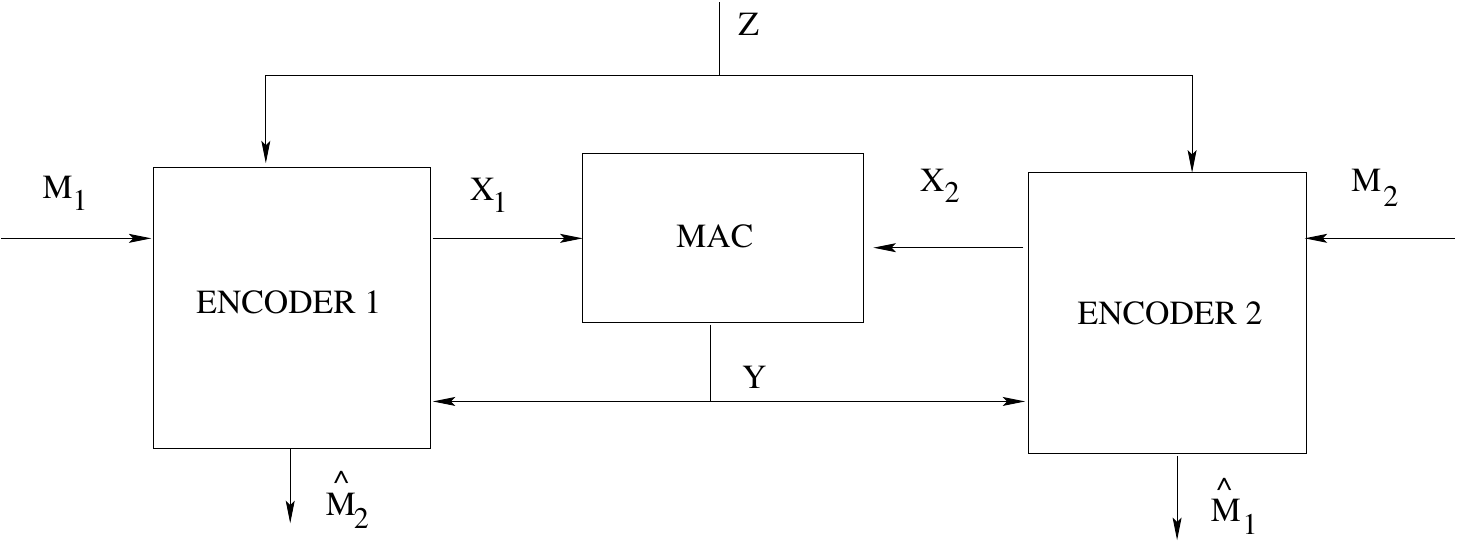}
\caption{Second phase: transmission of correlated information with
  common side information $Z$ on common output two-way channel. $Z$ is the
  channel output of phase one.}
\label{fig:phase_second}
\end{figure}

\emph{Second Phase}: The situation at the end of the first phase is as if a random edge is picked  from the above message graph with encoder $1$ knowing just the left vertex of  this edge, and encoder $2$ knowing just the  right vertex.  The two  encoders now have to  communicate over the channel so that each of them can recover this edge. The channel output block of the previous phase can be thought of as common side  information observed by all terminals.  This second phase of communication can be thought of as two terminals transmitting correlated information over a common  output two-way channel with common side information,
as shown in Figure \ref{fig:phase_second}. We note that the common side-information is `source state' rather than `channel state'- the output block of the previous phase is correlated with the messages (source of information) of the current phase. The channel behavior in the second phase does not depend on the common  side information since the channel is assumed to be
memoryless.

One approach to this communication problem is a strategy based on separate-source-channel coding: first perform distributed compression of the correlated messages (conditioned on the common side information) to produce two nearly independent indices, then transmit this pair of indices using a two-way channel code. This strategy of separate source and channel coding is not optimal in general. A more efficient way to transmit is to accomplish  this jointly: each encoder
 maps its message and the side information directly to the  channel input. By doing this, the two encoders recover the messages of each other at the end of the second phase. In other words, conditioned on the channel output blocks  of the two phases, the messages of the two encoders become perfectly
 correlated with high probability. The decoder however still  cannot recover these messages and has a list of highly likely message pairs.

\begin{figure}
\centering
\includegraphics[width=3.4in]{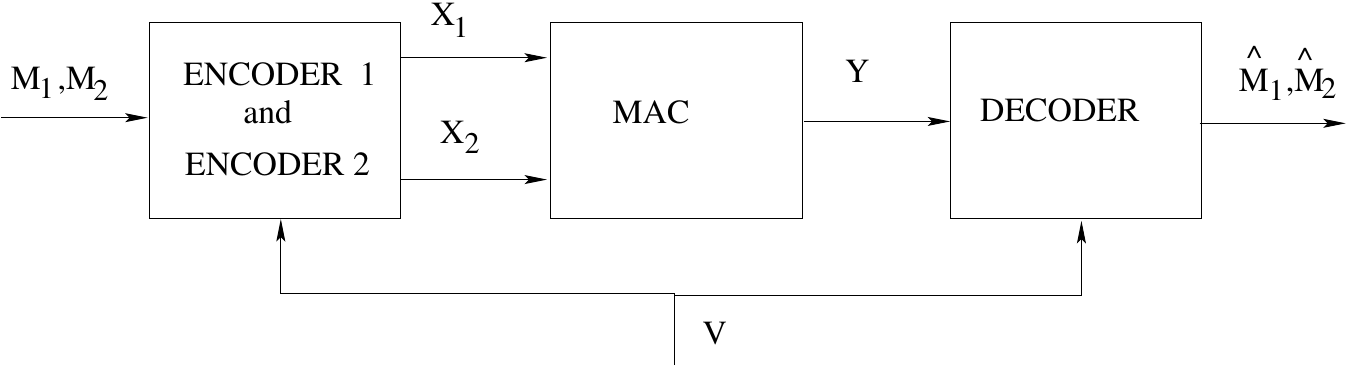}
\caption{Third phase: transmission of information with
  common side information $V$ on point-to-point channel. $V$ is the
  channel output of phases one and two.}
\label{fig:phase_third}
\end{figure}

\emph{Third Phase}: In the final phase of communication, the encoders wish to send a common message over the channel to the decoder so that its list of highly likely message pairs is disambiguated. This is shown in Figure \ref{fig:phase_third}. This phase  can be thought of as transmission of a message over a
point-to-point channel by an encoder to a decoder, with both terminals having common side information (the channel output blocks of the previous two phases) that is statistically correlated with the message. As before, the channel behavior in this phase is independent of this side information owing to the memoryless nature of the channel. For this phase, separate source and channel coding is optimal.

Having gone through the basic idea, let us consider some of the issues involved in obtaining a single-letter characterization of the performance of such a system. Suppose one uses a random coding procedure for the first phase based on single-letter product distributions on the channel inputs. Then the message graph obtained at the end of this phase is a random subset of the conditionally jointly typical set of channel inputs given the channel output. Due to the law of large numbers, with high probability, this  message graph is nearly semi-regular \cite{stanley}, i.e., the degrees of the left vertices are nearly equal, and the degrees of the right vertices are nearly equal.

Transmission of correlated sources  and correlated message graphs over the MAC has been studied in \cite{CES80} and \cite{PradhanChoi07}, respectively. In the former, the correlated  information is modeled as a pair of  memoryless correlated sources with a single-letter joint probability distribution. Unlike the model in \cite{CES80}, the statistical correlation of the  messages at the beginning of the second phase  cannot be captured by a single-letter probability distribution; rather, the correlation is captured by a message graph that is a random subset of a conditionally typical set. In other words, the random edges in the message graph do not exhibit a memoryless-source-like behavior.

In \cite{PradhanChoi07}, the correlation of the messages is modeled as a sequence of random edges from a sequence of nearly semi-regular bipartite graphs with increasing size. Inspired by the approaches of both \cite{CES80} and \cite{PradhanChoi07}, for the two-way communication in the second phase,
we will construct a joint-source-channel coding scheme that takes advantage of the common side information.

At the beginning of the third phase, the uncertainty list of the decoder consists of the likely message pairs conditioned on the channel
outputs of the previous two blocks. Due to the law of large numbers, each message pair in this list is nearly equally likely  to be the one  transmitted by the encoders in the first phase. This leads to a simple coding strategy for the third phase: a one-to-one mapping that maps the message pairs in the list to an index set, followed by channel coding to transmit the index over a point-to-point channel.

Finally, we superimpose the three phases to obtain a  new block-Markov superposition coding scheme.  Fresh information enters in each block and is resolved over the next two  blocks.  This scheme dictates the joint distributions we may choose for coding.

It turns out that there is one more hurdle to cross before we obtain a single-letter characterization - we need to ensure the stationarity
of the coding scheme.  Recall that in the second phase, each encoder generates its channel input based on its own message and the common side information.
 The channel inputs of the two encoders are correlated, and we need the joint distribution of these correlated inputs to be the same in each block.
 We ensure this by imposing a condition on the distributions used at the encoders to generate these correlated channel inputs. This leads to stationarity, resulting in a single-letter characterization. We show that  this  scheme yields a single-letter rate region involving three auxiliary random variables that includes the C-L region, and that the inclusion is strict using two examples.

Looking back, we make a couple of comments. At the beginning of the first phase, it is easy to see that the independent messages of the encoders can be thought of as a random edge in a \emph{fully connected} bipartite graph. In other words, since each pair of messages is equally likely to be transmitted in the first phase, every left vertex in the message graph is connected to every right vertex. The message graph gets progressively thinner over the three phases, until (with high probability) it reduces to a single edge at the end of the third phase. We note that this thinning of the message graph could be accomplished in four phases or even more. This results in improved rate regions involving a larger collection of  auxiliary random variables.

In the rest of the paper, we shall consider a formal treatment of the problem.  
In Section \ref{sec:main}, we give the required definitions and state the main result of the paper. In Section \ref{sec:coding_scheme}, we use bipartite message graphs to explain the main ideas behind the coding scheme quantitatively. In Section \ref{sec:comparisons}, we compare the proposed region with others in the literature using a couple of examples. The formal proof of the main theorem is given in Section \ref{sec:proof}. In Section \ref{sec:three_stage},
we show how our coding scheme can be extended to obtain larger rate regions with additional auxiliary random variables. Section \ref{sec:conclusion} concludes the paper.

\emph{Notation}: We use uppercase letters to denote random variables, lower-case for their realizations and bold-face notation for random vectors. Unless otherwise stated, all vectors have length $N$. Thus $\mathbf{A} \triangleq A^N \triangleq (A_1,\ldots,A_N)$. For any $\alpha$ such that $0<\alpha<1$, $\bar{\alpha}\triangleq 1-\alpha$. Unless otherwise mentioned, logarithms are with base $2$, and entropy and mutual information are measured in bits.

\section{Preliminaries and Main Result} \label{sec:main}
A two-user discrete memoryless MAC is defined by a quadruple $(\mathcal{X}_1,\mathcal{X}_2,\mathcal{Y},P_{Y|X_1,X_2})$ of
 input alphabets $\mathcal{X}_1,\mathcal{X}_2$ and output alphabet $\mathcal{Y}$, and a  set of probability
 distributions $P_{Y|X_1X_2}(.|x_1,x_2)$ on $\mathcal{Y}$ for all $x_1 \in \mathcal{X}_1, x_2 \in \mathcal{X}_2$.
The channel law for $n$ channel uses satisfies the following for all $n=1,2,\ldots$
\begin{equation*}
\begin{split}
& \mbox{Pr}(Y_n=y_n|X_1^n=\mathbf{x}_1,X_2^n=\mathbf{x}_2,Y^{n-1}=\mathbf{y})\\
& =P_{Y|X_1X_2}(y_n|x_{1n},x_{2n})
\end{split}
\end{equation*}
for all $y_n \in \mathcal{Y}$, $\mathbf{x}_1 \in \mathcal{X}_1^n$,
$\mathbf{x}_2 \in \mathcal{X}_1^n$ and $\mathbf{y} \in
\mathcal{Y}^{n-1}$.
There is noiseless feedback to both encoders ($S_1$ and $S_2$ are both closed in Figure \ref{fig:macfb}).
\begin{defi}
An $(N,M_1,M_2)$  transmission system  for a given MAC with feedback  consists of
\begin{enumerate}
\item A sequence of mappings for each encoder:
\ben
\begin{split}
e_{1n}:\{1,\ldots,M_1\} & \times \mathcal{Y}^{n-1} \to
{\mathcal{X}}_1, \quad n=1,\ldots,N \\
e_{2n}:\{1,\ldots,M_2\} & \times \mathcal{Y}^{n-1} \to
{\mathcal{X}}_2, \quad n=1,\ldots,N
\end{split}
\een
\item A decoder mapping  given by
\ben
g:  \mathcal{Y}^{N} \to \{1,\ldots,M_1\} \times \{1,\ldots,M_2\}  \\
\een
\end{enumerate}
\end{defi}
We assume that  the messages $(W_1,W_2)$ are drawn uniformly from the set $\{1,\ldots,M_1\} \times\{1,\ldots,M_2\}$. The channel input
of encoder $i$ at time $n$ is given by $X_{in}=e_{in}(W_i,Y^{n-1})$ for $n=1,2,\ldots,N$ and $i=1,2$.
The average error probability of the above transmission system is given by
{\small{
\ben
\tau= \frac{1}{M_1M_2}
\sum_{w_1=1}^{M_1} \sum_{w_2=1}^{M_2}
\text{Pr}(g(\mathbf{Y}) \neq (w_1,w_2)| W_1,W_2=w_1,w_2).
\een
}}
\begin{defi}
A rate pair $(R_1,R_2)$ is said to be achievable for a given discrete  memoryless MAC with feedback if $\forall \epsilon>0$,
there exists an $N(\epsilon)$ such that for all $N> N(\epsilon)$ there exists an $(N,M_1,M_2)$ transmission systems that satisfies the following
conditions
\[
\frac{1}{N} \log M_1 \geq R_1- \epsilon,  \ \
\ \ \frac{1}{N} \log M_2 \geq R_2-\epsilon, \ \
\tau \leq \epsilon.
\]
The set of all achievable rate pairs is the capacity region with feedback.
\end{defi}
The following theorem is the main result of this paper.
\begin{defi} \label{def:consistency}
For a given MAC
$(\mathcal{X}_1,\mathcal{X}_2,\mathcal{Y},P_{Y|X_1,X_2})$ define  $\mathcal{P}$ as the set of all distributions  $P$ on
$\mathcal{U} \times \mathcal{A} \times \mathcal{B} \times \mathcal{X}_1 \times \mathcal{X}_2 \times \mathcal{Y}$ of the form
\be
P_UP_{AB}P_{X_1|UA}P_{X_2|UB}P_{Y|X_1X_2}
\label{eq:joint_dist1}
\ee
where $\mathcal{U},\mathcal{A}$ and  $\mathcal{B}$ are arbitrary finite sets. Consider two sets of random variables $(U,A,B,X_1,X_2,Y)$ and
$(\tilde{U},\tilde{A}, \tilde{B},\tilde{X}_1,\tilde{X}_2,\tilde{Y})$ each having the above distribution $P$. For conciseness, we often refer to the collection $(U,A,B,Y)$ as $S$, $(\tilde{U},\tilde{A},\tilde{B},\tilde{Y})$ as $\tilde{S}$, and $\mathcal{U} \times \mathcal{A} \times \mathcal{B} \times \mathcal{Y}$ as $\mathcal{S}$. Hence
\[
P_{SX_1X_2}=P_{\tilde{S}\tilde{X}_1\tilde{X}_2}=P.
\]
Define $\mathcal{Q}$ as the set of  pairs of  conditional distributions $(Q_{A|\tilde{S},\tilde{X}_1},Q_{B|\tilde{S},\tilde{X}_2})$,
that satisfy  the following consistency condition
{\small{
\begin{equation} \label{eq:q1q2cond}
\begin{split}
& \sum_{\tilde{s},\tilde{x}_1,\tilde{x}_2 \in \mathcal{S} \times \mathcal{X}_1 \times \mathcal{X}_2}
\hspace{-15pt} P_{\tilde{S} \tilde{X}_1 \tilde{X}_2}(\tilde{s},\tilde{x}_1,\tilde{x}_2)
Q_{A|\tilde{S},\tilde{X}_1}(a|\tilde{s},\tilde{x}_1) Q_{B|\tilde{S},\tilde{X}_2}(b|\tilde{s},\tilde{x}_2)\\
&=P_{AB}(a,b), \quad \forall (a,b) \in \mathcal{A} \times \mathcal{B}.
\end{split}
\end{equation}
}}
Then,  for any  $(Q_{A|\tilde{S},\tilde{X}_1},Q_{B|\tilde{S},\tilde{X}_2}) \in \mathcal{Q}$,
the joint distribution of the two sets of random variables - $(\tilde{S},\tilde{X}_1,\tilde{X}_2)$ and
$({S},{X}_1,{X}_2)$ -  is
given by
\be \label{eq:two_block_dist}
P_{\tilde{S} \tilde{X}_1 \tilde{X}_2}
Q_{A|\tilde{S},\tilde{X}_1}
Q_{B|\tilde{S},\tilde{X}_2}
P_{UX_1X_2Y|AB}.
\ee
\end{defi}

\begin{thm} \label{thm:macfb_better}
For a MAC $(\mathcal{X}_1,\mathcal{X}_2,\mathcal{Y},P_{Y|X_1,X_2})$,
for any distribution $P$ from $\mathcal{P}$ and a pair of
conditional distributions $(Q_{A|\tilde{S},\tilde{X}_1}, Q_{B|\tilde{S},\tilde{X}_2})$ from $\mathcal{Q}$,
the following rate-region is achievable.
\be \label{eq:thm_statement}
\begin{split}
&R_1  \leq I(X_1;Y|X_2BU\tilde{S}\tilde{X}_2)\\
& \qquad -\left( I(A;X_2|YBU\tilde{S}\tilde{X}_2)-I(U;Y|\tilde U \tilde Y) \right)^+ , \\
&R_2 \leq I(X_2;Y|X_1AU \tilde{S}\tilde{X}_1)\\
& \qquad -\left( I(B;X_1|YAU \tilde{S} \tilde{X}_1)-I(U;Y|\tilde U \tilde Y) \right)^+ , \\
&R_1+R_2 \leq  I(X_1X_2;Y| U \tilde S) + I(U;Y|\tilde U \tilde Y).
\end{split}
\ee
\end{thm}
In the above, we have used $x^+$ to denote $\max(0,x)$.
If we set $A=B=\phi$, we obtain the Cover-Leung region, specified by
\eqref{eq:CL-1}-\eqref{eq:CL-3}  in next section.

\emph{Remark}: The rate region of Theorem \ref{thm:macfb_better} is convex.

\section{The Coding Scheme}\label{sec:coding_scheme}
In this  section, we give a sketch of the proof of the coding theorem. The discussion here is informal; the formal proof of the theorem is given in Section \ref{sec:proof}. As we have seen in Section \ref{sec:intro}, to visualize the ideas behind the coding scheme, it is  useful to represent the messages of the two encoders in terms of a bipartite graph. Let us suppose that the two encoders wish to transmit independent information at rates $R_1$ and $R_2$, respectively and
use the channel $N$ times. Before transmission begins, the message graph is a fully connected bipartite graph with $2^{NR_1}$ left vertices and $2^{NR_2}$ right vertices. This graph is shown in Figure \ref{fig:clgraph}(a), where each left vertex denotes a message of encoder $1$ and each right vertex represents a
message of encoder $2$. An edge connecting two vertices represents a message pair that has non-zero probability.

We shall first review the C-L scheme in the framework of message graphs, and then extend the ideas to develop our coding scheme.

\subsection{The Cover-Leung Scheme}
\begin{figure}
\centering
\begin{tikzpicture}[scale=0.8]
\foreach \x in {2,...,10}
{
\node[leftvertex] (L\x) at (-0.5,\x)  {};
\foreach \y in {2,...,10}
    {
    \node[rightvertex] (R\y) at (3,\y)  {};
    \draw [-] (L\x) to (R\y);
    }
}
\draw [-] (L2) to node[below=5pt] {(a)} (R2);
\foreach \x in {11,...,7}
{
\node[leftvertex] (L\x) at (5,\x)  {};
\node[rightvertex] (R\x) at (8,\x)  {};
}
\draw [-] (L11) to (R9); \draw [-] (L10) to (R8);
\draw [-] (L9) to (R11);\draw [-] (L8) to node[below=15pt] {(b)}(R7);
\draw [-] (L7) to (R10); 
\foreach \x in {5,4,...,0}
{
\node[leftvertex] (L\x) at (5,\x)  {};
\node[rightvertex] (R\x) at (8,\x)  {};
\draw [-] (L\x) to (R\x);
}
\draw (L0) to node[below=5pt] {(c)} (R0);
\draw [-] (L0) to (R1); \draw [-] (L0) to (R5);
\draw [-] (L1) to (R3); \draw [-] (L1) to (R0);
\draw [-] (L2) to (R3); \draw [-] (L2) to (R1);
\draw [-] (L3) to (R4); \draw [-] (L3) to (R2);
\draw [-] (L4) to (R5); \draw [-] (L4) to (R3);
\draw [-] (L5) to (R0); \draw [-] (L5) to (R4);
\end{tikzpicture}
\caption{Decoder's message graph for the C-L scheme: (a) Before transmission (b) When each encoder can decode the other's message upon 
receiving block output $\mathbf{Y}$ (c) When the encoders cannot decode  from the output $\mathbf{Y}$ }
\label{fig:clgraph}
\end{figure}
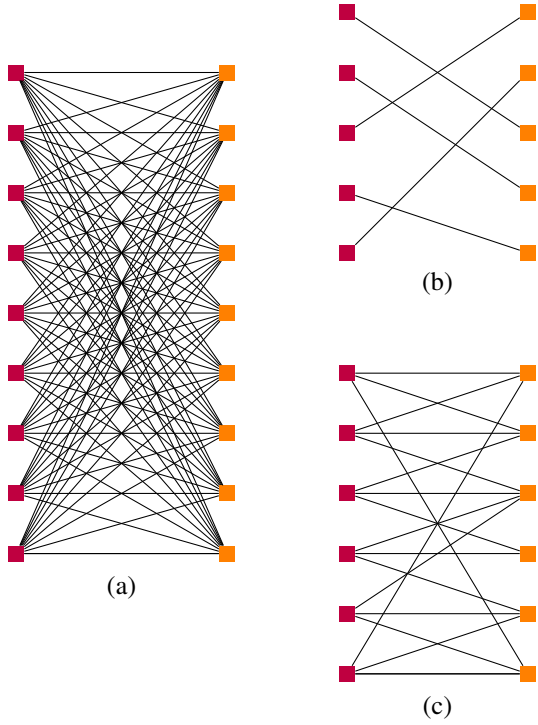

\begin{fact} \emph{Cover-Leung (C-L) Region \cite{CL81}}:
Consider a joint distribution  of the form
$P_{UX_1X_2Y}$ $=P_{U}P_{X_1|U}$ $P_{X_2|U}P_{Y|X_1X_2}$,
where $P_{Y|X_1X_2}$ is fixed by the channel and  $U$ is a discrete random variable with cardinality
$\min\{|\mathcal{X}_1|\cdot|\mathcal{X}_2|+1, |\mathcal{Y}|+2\}$. Then the
following rate pairs $(R_1,R_2)$ are achievable.
\begin{gather}
R_1  < I(X_1;Y|X_2U), \label{eq:CL-1}\\
R_2 < I(X_2;Y|X_1U), \label{eq:CL-2}\\
R_1+R_2 < I(X_1 X_2;Y) \label{eq:CL-3}.
\end{gather}
\end{fact}

In this scheme, there are $L$ blocks of transmission, with a fresh pair of messages in each block. Let $(W_{1l},W_{2l}), 1\leq l < L$, denote the message pair for block $l$, drawn from sets of size $2^{NR_1}$ and $2^{NR_2}$, respectively. The codebooks of the two encoders for each block are drawn i.i.d according to distributions $P_{X_1|U}$ and $P_{X_2|U}$, respectively, where $U$ is an auxiliary random variable known to both transmitters. Let $(\mathbf{X}_{1l},\mathbf{X}_{2l})$ denote the codewords corresponding to the message pair. $(W_{1l},W_{2l})$ (or equivalently, $(\mathbf{X}_{1l},\mathbf{X}_{2l})$) corresponds to a random edge in the graph of  Figure \ref{fig:clgraph}(a). After the decoder receives the output $\mathbf{Y}_l$, the  message graph conditioned on the channel output (posterior message graph) for block $l$ is the set of all message pairs $(W_{1l},W_{2l})$ that could have occurred given $\mathbf{Y}_l$. We can define a high probability subset of the posterior message graph, which we call the \emph{effective} posterior message graph, as follows. Let  $\mathcal{L}_l$ be the set of all message pairs $(i,j)$ such that $(\mathbf{X}_{1l}(i),\mathbf{X}_{2l}(j), \mathbf{Y}_l)$ are jointly typical. The edges of the effective posterior message graph  are the message pairs contained in $\mathcal{L}_l$.

If the rate pair $(R_1,R_2)$ lies outside the no-feedback capacity region, the decoder cannot decode $(W_{1l},W_{2l})$ from the output
$\mathbf{Y}_l$. Owing to feedback,  both encoders know $\mathbf{Y}_l$ at the end of block $l$. If $R_1$ and $R_2$ satisfy \eqref{eq:CL-1} and
\eqref{eq:CL-2}, it can be shown that  using the feedback, each encoder can correctly decode the message of the other with high probability. In other words,
each edge of the effective posterior message graph is uniquely determined by knowing either the left vertex
or the right vertex. Thus, upon receiving $\mathbf{Y}_l$,  the effective posterior message graph \emph{at the decoder} has the structure shown in Figure
\ref{fig:clgraph}(b). The number of edges in this graph is approximately
 \[ 2^{N(R_1+R_2-I(X_1X_2;Y|U))}. \]
The two encoders cooperate to resolve this decoder uncertainty using a common codebook of $\mathbf{U}$ sequences.
This codebook has size $2^{NR_0}$, with each codeword symbol chosen i.i.d according to $P_U$. Each codeword indexes
 an edge in the message graph of Figure \ref{fig:clgraph}(b).  Since both encoders know the random edge $(W_{1l},W_{2l})$,
 they pick the appropriate codeword from this codebook and set it as $\mathbf{U}_{l+1}$. $\mathbf{U}_{l+1}$ uniquely
 specifies the edge in the graph  if the codebook size is greater than the number of edges in the graph of
Figure \ref{fig:clgraph}(b). This happens if
\be \label{eq:CL-dec1} R_0 >  R_1+R_2-I(X_1X_2;Y|U). \ee

 The codewords $\mathbf{X}_{1(l+1)}, \mathbf{X}_{2(l+1)}$ carry fresh messages for block $(l+1)$, and
 are picked conditioned on $\mathbf{U}_{l+1}$ according to $P_{X_1|U}$ and $P_{X_2|U}$, respectively. Thus in each block,
 fresh information is superimposed on resolution information for the previous block.
 The decoder can decode $\mathbf{U}_{l+1}$ from $\mathbf{Y}_{l+1}$ if the rate $R_0$ of the $U$-codebook satisfies
 \be  \label{eq:CL-dec2}
 R_0 < I(U;Y)
 \ee
Combining \eqref{eq:CL-dec1} and \eqref{eq:CL-dec2}, we obtain the final constraint \eqref{eq:CL-3} of the C-L rate region.

\subsection{Proposed Coding scheme}
Suppose that the rate pair $(R_1,R_2)$ lies outside the C-L region. Then at the end of each block $l$, the encoders cannot decode the message of
one another. The effective posterior message graph at the decoder on receiving $\mathbf{Y}_l$ now looks like Figure \ref{fig:clgraph}(c) -
with high probability, each vertex no longer has degree one. The degree of each left vertex $\mathbf{X}_{1l}$ is the number of codewords $\mathbf{X}_{2l}$ that are jointly typical with $(\mathbf{X}_{1l},\mathbf{Y}_l)$. This number is  approximately $2^{N(R_2-I(X_2;Y|X_1U))}$. Similarly, the degree of each right vertex
is approximately $2^{N(R_1-I(X_1;Y|X_2U))}$. The number of left vertices is approximately $2^{N(R_1-I(X_1;Y))}$ and the number of right vertices is approximately $2^{N(R_2-I(X_2;Y))}$. This graph is nearly semi-regular. Moreover, since the channel output is a random sequence, this graph is a random subset of the conditionally typical set of $(X_1,X_2)$ given $(Y,U)$.

Clearly, the uncertainty of the decoder about $(W_{1l},W_{2l})$ now cannot be resolved with just a common message since both encoders cannot agree on the edge in the effective posterior message graph. Of course, conditioned on $\mathbf{Y}_l$, the messages are \emph{correlated}, rather than independent. In other words, the effective posterior message graph conditioned on $\mathbf{Y}_l$ in Figure \ref{fig:clgraph}(c) has left and right degrees that are strictly less than $R_1$ and $R_2$, respectively. The objective now is to efficiently transmit the random edge $(W_{1l},W_{2l})$ from the effective message graph of Figure \ref{fig:clgraph}(c).

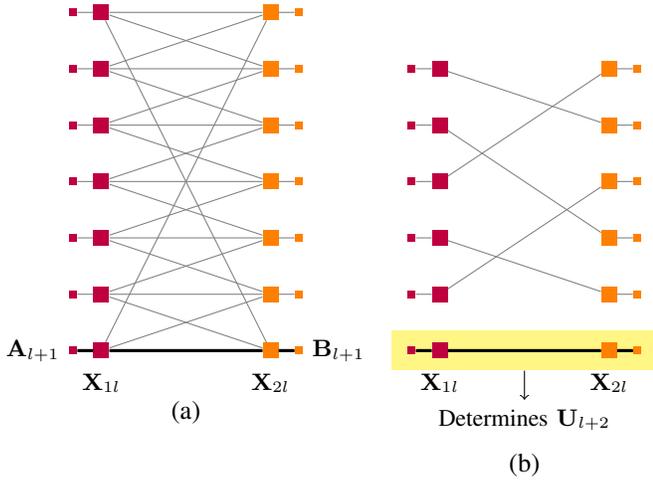
\begin{figure}
\centering
\begin{tikzpicture}[scale=0.75]
\foreach \x in {8,7,...,2}
{
\node[leftvertex] (L\x) at (0,\x)  {};
\node[Sleftvertex] (LA\x) at (-0.5,\x) {};
\node[rightvertex] (R\x) at (3,\x)  {};
\node[Srightvertex] (RB\x) at (3.5,\x) {};
\draw [-, thin, gray] (L\x) to (R\x);
\draw [-, thin, gray] (L\x) to (LA\x);
\draw [-, thin, gray] (R\x) to (RB\x);
}
\draw [-, very thick] (L2) to node[below=15pt] {(a)} (R2);
\draw [-, very thick] (L2) to (LA2);
\draw [-, very thick] (R2) to (RB2);
\draw [-, thin, gray] (L2) to (R3); \draw [-, thin, gray] (L2) to (R8);
\draw [-, thin, gray] (L3) to (R4); \draw [-, thin, gray] (L3) to (R2);
\draw [-, thin, gray] (L4) to (R5); \draw [-, thin, gray] (L4) to (R3);
\draw [-, thin, gray] (L5) to (R6); \draw [-, thin, gray] (L5) to (R4);
\draw [-, thin, gray] (L6) to (R7); \draw [-, thin, gray] (L6) to (R5);
\draw [-, thin, gray] (L7) to (R8); \draw [-, thin, gray] (L7) to (R6);
\draw [-, thin, gray] (L8) to (R2); \draw [-, thin, gray] (L8) to (R7);
\node [below=2pt] at (L2.south) {\small{$\mathbf{X}_{1l}$}};
\node [below=2pt] at (R2.south) {\small{$\mathbf{X}_{2l}$}};
\node [left] at (LA2.west) {\small{$\mathbf{A}_{l+1}$}};
\node [right] at (RB2.east) {\small{$\mathbf{B}_{l+1}$}};
\foreach \x in {7,...,2}
{
\node[leftvertex] (L\x) at (6,\x)  {};
\node[Sleftvertex] (LA\x) at (5.5,\x) {};
\node[rightvertex] (R\x) at (9,\x)  {};
\node[Srightvertex] (RB\x) at (9.5,\x) {};
\draw [-, thin, gray] (L\x) to (LA\x);
\draw [-, thin, gray] (R\x) to (RB\x);
}
\draw [-, thin, gray] (L7) to (R6); \draw [-, thin, gray] (L6) to (R4);
\draw [-, thin, gray] (L5) to (R7);\draw [-, thin, gray] (L4) to (R3);
\draw [-, thin, gray] (L3) to (R5); \draw [-, thin, gray] (L2) to (R2);
\draw [-, very thick] (L2) to node[below=35pt] {(b)} (R2);
\draw [-, very thick] (L2) to (LA2);
\draw [-, very thick] (R2) to (RB2);
\node [below=2pt] at (L2.south) {\small{$\mathbf{X}_{1l}$}};
\node [below=2pt] at (R2.south) {\small{$\mathbf{X}_{2l}$}};
\begin{pgfonlayer}{background}
\node [fill=yellow!60, fit=(L2) (R2)] (BG) {};
\end{pgfonlayer}
\node [below=11pt] (txt) at (BG.south) {\small{Determines $\mathbf{U}_{l+2}$}};
\draw [->] (BG) -- (txt);
\end{tikzpicture}
\caption{Message graph for the pair $(W_{1l},W_{2l})$, the transmitted message pair shown in bold-face:  a) After receiving $\mathbf{Y}_{l}$ b) After receiving $\mathbf{Y}_{l+1}$}
\label{fig:three_stage}
\end{figure}

 Generate a sequence $\mathbf{A}$ for each jointly typical sequence pair $(\mathbf{X_1,Y})$, with symbols generated i.i.d from the distribution $P_{A|X_1 Y}$. Similarly, generate a sequence  $\mathbf{B}$ for each jointly typical pair $(\mathbf{X_2,Y})$, according to  distribution $P_{B|X_2 Y}$.
Recall that $(\mathbf{X}_{1l},\mathbf{X}_{2l})$ denotes the codeword pair transmitted in block $l$. Encoder $1$ sets $\mathbf{A}_{l+1}$ equal to
the $A$-sequence corresponding to $(\mathbf{X}_{1l},\mathbf{Y}_l)$, and encoder $2$ sets $\mathbf{B}_{l+1}$ equal to the
$B$-sequence corresponding to $(\mathbf{X}_{2l},\mathbf{Y}_l)$. This is shown in Figure \ref{fig:three_stage}(a).
The codeword $\mathbf{X}_{1(l+1)}$, which carries a fresh message for block $(l+1)$, is chosen conditioned on $\mathbf{A}_{l+1}$. Similarly,
$\mathbf{X}_{2(l+1)}$ is chosen conditioned on $\mathbf{B}_{l+1}$. We note that $\mathbf{A}_{l+1}$ and $\mathbf{B}_{l+1}$ are
correlated since they are chosen conditioned on $(\mathbf{X}_{1l},\mathbf{Y}_l)$ and $(\mathbf{X}_{2l},\mathbf{Y}_l)$, respectively.

At the end of block $(l+1)$,  the decoder and the two encoders receive $\mathbf{Y}_{l+1}$. Encoder $1$  decodes $\mathbf{B}_{l+1}$ from
$(\mathbf{Y}_{l+1},\mathbf{A}_{l+1}, \mathbf{X}_{1l})$. Similarly, encoder $2$ decodes $\mathbf{A}_{l+1}$ from $(\mathbf{Y}_{l+1},\mathbf{B}_{l+1}, \mathbf{X}_{2l})$. Assuming this is done correctly, both encoders now know the message pair $(W_{1l},W_{2l})$, but the decoder does not, since it may not be able decode $(\mathbf{A}_{l+1},\mathbf{B}_{l+1})$ from $\mathbf{Y}_{l+1}$. Then the effective posterior message graph at the decoder on receiving $\mathbf{Y}_{l+1}$
 has the form shown in Figure \ref{fig:three_stage}(b). Since both encoders now know the edge in the effective posterior message graph conditioned on $(\mathbf{Y}_l, \mathbf{Y}_{l+1})$ corresponding to $(W_{1l},W_{2l})$, they can cooperate to resolve the decoder's uncertainty using a common sequence $\mathbf{U}_{l+2}$
in block $(l+2)$.

To summarize, codewords $(\mathbf{X}_{1l},\mathbf{X}_{2l})$ which carry the fresh messages for block $l$, can be decoded by neither the encoders nor the decoder upon receiving $\mathbf{Y}_l$. So the encoders send correlated information $(\mathbf{A}_{l+1},\mathbf{B}_{l+1})$ in block $(l+1)$ to help each other decode $(W_{1l},W_{2l})$. They then cooperate to send $\mathbf{U}_{l+2}$, so that the decoder can decode $(W_{1l},W_{2l})$ at the end of block $(l+2)$. In the `one-step' C-L coding scheme, the rates $(R_1,R_2)$ are low enough so that each encoder can decode the message of the other at the end of the {same} block.   In other words, the fully-connected graph of Figure \ref{fig:clgraph}(a) is thinned to the degree-$1$ graph of Figure \ref{fig:clgraph}(b) in one block.
In our `two-step' strategy, the thinning of the fully-connected graph to the degree-1 graph takes place over \emph{two} blocks as shown in Figure \ref{fig:three_stage}.

\begin{figure}
\centering
\includegraphics[height=2in]{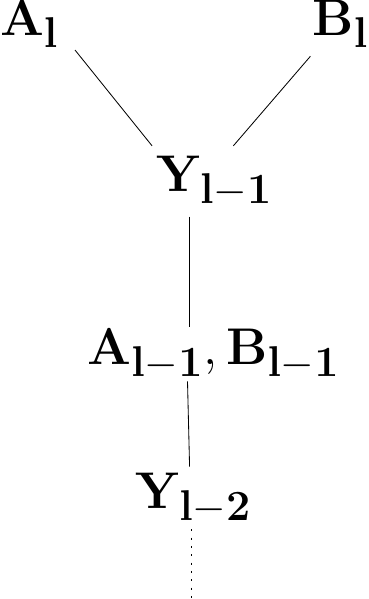}
\caption{Correlation propagates across blocks}
\label{fig:depend}
\end{figure}

\subsubsection{Stationarity of the coding scheme}
 The scheme  proposed above has a shortcoming - it is not stationary  and hence does not yield a single-letter rate region.
 Recall that for any block $l$, $\mathbf{A}_{l}$ and $\mathbf{B}_{l}$  are produced conditioned on $\mathbf{Y}_{l-1}$. $\mathbf{Y}_{l-1}$ is
 produced by the channel based on inputs $(\mathbf{X}_{1(l-1)},\mathbf{X}_{2(l-1)})$, which in turn depend on
 $\mathbf{A}_{l-1}$ and $\mathbf{B}_{l-1}$, respectively. Thus we have  correlation that propagates across blocks, as shown in  Figure
 \ref{fig:depend}. This implies that the resulting rate region will be a  multi-letter characterization  that depends on the joint distribution
 of the variables in \emph{all $L$ blocks}: $\left\{ \left(U_l,A_l,B_l, X_{1l},X_{2l}, Y_l\right)\right\}_{l=1}^L$.

  To obtain a single-letter rate region, we require a stationary  distribution of sequences in each block.
In other words, we need the random sequences  $(\mathbf{U},\mathbf{A},\mathbf{B},\mathbf{X}_1\mathbf{X}_2,\mathbf{Y})$ to be characterized
by the same single-letter product distribution in each block. This will happen if we can ensure that the
$\mathbf{A},\mathbf{B}$ sequences in each block have the same single-letter distribution  $P_{AB}$.
  The correlation between $\mathbf{A}_{l+1}$ and $\mathbf{B}_{l+1}$ cannot be arbitrary - it is
  generated using the information available at each encoder at the end of block $l$. At this time,
  both encoders know $\mathbf{s}_l \triangleq (\mathbf{u,a,b,y})_{l}$. In addition, encoder  $1$ also knows $\mathbf{x}_{1l}$
  and hence we make it generate $\mathbf{A}_{l+1}$ according to the product distribution
  $Q_{A|\tilde S \tilde X_1}^n(.|\mathbf{s}_{l},\mathbf{x}_{1l})$. Similarly, we make encoder $2$ generate  $\mathbf{B}_{l+1}$
  according to $Q_{B|\tilde S \tilde X_2}^n(.|\mathbf{s}_{l},\mathbf{x}_{2l})$. If the pair
  $(Q_{A|\tilde S \tilde X_1},Q_{B|\tilde S \tilde X_2})\in \mathcal{Q}$, then equation \eqref{eq:q1q2cond} ensures that
  the pair $(\mathbf{A}_{l+1},\mathbf{B}_{l+1})$  corresponding to $(W_{1l},W_{2l})$ belongs to the typical set  $T(P_{AB})$
  with high probability. This ensures stationarity of  the coding scheme.

Our block-Markov coding scheme, with conditions imposed to ensure stationarity, is similar in spirit to that of Han for two-way channels \cite{Han84}.
Finally, a couple of comments on the chosen input distribution in \eqref{eq:joint_dist1}. In block $(l+1)$, the encoders generate $\mathbf{A}_{l+1}$ and $\mathbf{B}_{l+1}$ independently based on their own messages for block $l$ and the common side information $\mathbf{S}_l = (\mathbf{U,A,B,Y})_{l}$. Why do they not use $\mathbf{S}_{l-1}, \mathbf{S}_{l-2}, \ldots$ (the side information accumulated from earlier blocks) as well? This is because $(W_{1(l-2)},W_{2(l-2)})$ is decoded at the decoder at the end of block $l$, and $(W_{1(l-2)},W_{2(l-2)})$ determines $(\mathbf{A,B})_{l-1}$. Hence, for block $(l+1)$, $\mathbf{S}_{l-1}, \mathbf{S}_{l-2}, \ldots$ is known at all terminals and is just
shared common randomness.

Also note that $U$, which carries common information sent by both encoders, is independent of the random variables $(A,B)$. It is sufficient to choose a distribution of the form $P_UP_{AB}$ (rather than $P_{UAB}$). This is because separate source and channel coding is optimal when the encoders send common information over the MAC. Joint source-channel coding is needed only for sending correlated information. Hence $\mathbf{A},\mathbf{B}$ are generated conditioned on the information available at each encoder, but $\mathbf{U}$ is generated independently.

We remark that our scheme can be extended as follows. The above coding scheme thins the fully-connected graph  to the degree-one graph over two blocks. Instead, we could do it over three blocks, going through two intermediate stages of progressively thinner (more correlated) graphs before obtaining the degree-one graph. This would yield a potentially larger rate region, albeit with extra  auxiliary random variables. This is discussed in Section  \ref{sec:three_stage}.

\section{Comparisons}\label{sec:comparisons}
In this section, the rate region of Theorem \ref{thm:macfb_better} is compared with the other known regions for the memoryless MAC with noiseless feedback. We first consider the white Gaussian MAC. Since its feedback capacity is known \cite{Ozarow84}, this channel provides a benchmark to compare the rate region of Theorem \ref{thm:macfb_better}. We see that our rate region yields rates strictly better than the C-L region, but smaller than the feedback capacity.
 Ozarow's capacity-achieving scheme in \cite{Ozarow84} is specific to the Gaussian case and does not extend to other MACs. The rate regions of Kramer \cite{Kramer98} and  Bross and Lapidoth (B-L)\cite{BrossLapi05} extend the C-L region for a discrete memoryless MAC with feedback. We compare our scheme with these in Sections \ref{subsec:kramer_comp} and \ref{subsec:bl_comp}.

We mention that all the calculations in this section are done using the rate constraints in \eqref{eq:achievable_region}, an equivalent representation of the rate constraints in Theorem \ref{thm:macfb_better}. This equivalence is established by equations \eqref{eq:final0v}-\eqref{eq:final2v} in Section \ref{sec:proof}.

\subsection{Additive White Gaussian MAC}\label{subsec:awgn}
Consider the AWGN MAC with power constraint $P$ on each of the inputs. This channel, with $\mathcal{X}_1=\mathcal{X}_2=\mathcal{Y}=\mathbb{R}$,
is defined by
\be \label{eq:channel} Y=X_1+X_2+N \ee
where $N$ is a Gaussian noise random variable with mean $0$ and variance $\sigma^2$ that is independent of $X_1$ and $X_2$.
The inputs $\mathbf{x}_1$ and $\mathbf{x}_2$ for each block satisfy $\frac{1}{N}\sum_{n=1}^N x_{1n}^2 \leq P,  \;  \frac{1}{N}\sum_{n=1}^N x_{2n}^2 \leq P.$
For this channel, the equal-rate point on the boundary of the C-L region \cite{CL81} is $(R_{CL},R_{CL})$
where
\be \label{eq:cl_eqpt}
R_{CL}=\frac{1}{2}\log\left(2\sqrt{1+\frac{P}{\sigma^2}}-1\right)
\ee

The achievable rate region of Theorem \ref{thm:macfb_better} for the discrete memoryless case can be extended to the AWGN MAC using a similar proof.
For the joint distribution $P_{UABX_1X_2Y}$ in \eqref{eq:joint_dist1}, define  $U \sim \mathcal{N}(0,1)$ and
$(A,B)$ jointly Gaussian with mean zero and covariance  matrix
\be \label{eq:kwab}
K_{AB}=
\begin{bmatrix}
1& \lambda\\
\lambda & 1
\end{bmatrix}.
\ee
The input distributions $P_{X_1|UA}$ and $P_{X_2|UB}$ are defined by
\be \label{eq:inp_dist}
\begin{split}
X_1&=\sqrt{\alpha P}\; I_{X_1} + \sqrt{\beta P}\; A + \sqrt{\overline{\alpha + \beta}P}\; U,\\
X_2&=\sqrt{\alpha P}\; I_{X_2} + \sqrt{\beta P}\; B + \sqrt{\overline{\alpha + \beta}P}\; U
\end{split}
\ee
where $I_{X_1}, I_{X_2}$ are independent $\mathcal{N}(0,1)$ random variables, $\alpha,\beta>0$ and
$\alpha+\beta \leq 1$. $I_{X_1}$ and $I_{X_2}$ represent the fresh information and $U$ is the resolution information for the decoder sent cooperatively by the encoders. $A$ and $B$ represent the information to be decoded using feedback by encoders $2$ and $1$, respectively.

Recall that $\tilde{S} \triangleq (\tilde{U},\tilde{A},\tilde{B},\tilde{Y})$. The distributions $Q_{A|\tilde{S} \tilde{X}_1 }$  and $Q_{B|\tilde{S} \tilde{X}_2}$ to generate $A$ and $B$ at the encoders are defined as
\be \label{eq:q1q2def}
\begin{split}
Q_{A|\tilde{S} \tilde{X}_1 }:  A = & k_1 \frac{\tilde X_1 - \sqrt{\overline{\alpha + \beta}P}\; \tilde U-\sqrt{\beta P}\; \tilde A}{\sqrt{\alpha P}} \\
&+ k_2 f(\tilde U, \tilde A, \tilde B,\tilde Y),\\
Q_{B|\tilde{S} \tilde{X}_2}: B=& -k_1 \frac{\tilde X_2 - \sqrt{\overline{\alpha + \beta}P}\; \tilde U -\sqrt{\beta P}\; \tilde B}{\sqrt{\alpha P}} \\
& - k_2 f(\tilde U,\tilde A,\tilde B,\tilde Y)
\end{split}
\ee
where $k_1,k_2 \in \mathbb{R}$ and
\be \label{eq:Wdef}
f(Y,A,B,U) \triangleq \frac{Y-\sqrt{\beta P}\;A-\sqrt{\beta P}\;B-2\sqrt{\overline{\alpha+\beta}P}\;U}{\sqrt{2\alpha P+\sigma^2}}.
\ee
It can be  verified that this choice of $(Q_{A|\tilde{S} \tilde{X}_1 }, Q_{B|\tilde{S} \tilde{X}_2})$ satisfies the consistency condition \eqref{eq:q1q2cond} (required for Theorem \ref{thm:macfb_better}) if the following equations are satisfied.
\be \label{eq:threeconds}
E[A^2]=E[B^2]=1, \; E[A B]=\lambda.
\ee
Using \eqref{eq:Wdef} and \eqref{eq:q1q2def}, the conditions in \eqref{eq:threeconds} become
\be \label{eq:ea2}
\begin{split}
1=E[{A}^2]=&k_1^2+k_2^2+2k_1k_2\sqrt{\frac{\alpha P}{2\alpha P+\sigma^2}},
\end{split}
\ee
\be \label{eq:eab}
\begin{split}
\lambda=E[{A}{B}]=-k_2^2-2k_1k_2\sqrt{\frac{\alpha P}{2\alpha P+\sigma^2}},
\end{split}
\ee
Adding \eqref{eq:ea2} and \eqref{eq:eab}, we get $k_1^2=1+\lambda.$
Substituting  $k_1 = \pm \sqrt{1+\lambda}$ in \eqref{eq:ea2} yields a quadratic equation
that can be solved to obtain $k_2$. The condition for the quadratic to yield a valid (real) solution for $k_2$ is
\be
\lambda \leq \frac{\alpha P}{\alpha P + \sigma^2}.
\ee

\begin{table}[t]
\caption{Comparison of equal-rate boundary points (in bits)}
\begin{center}
\vspace{-5pt}
\begin{tabular}{|c|c|c|c|c|c|}
\multicolumn{6}{c}{$ P/\sigma^2$}\\
\hline
&$0.5$&$1$&$5$&$10$&$100$ \\
\hline
$R_{CL}$ & $0.2678$ & $0.4353$ & $0.9815$ &$1.2470$& $2.1277$  \\
$R^{*}$ & $0.2753$ & $0.4499$ & $1.0067$ &$1.2709$& $2.1400$  \\
$R_{\text{FBcap}}$ & $0.2834$  & $0.4642$  & $1.0241$ & $1.2847$& $2.1439$ \\
\hline
\end{tabular}
\end{center}\label{tab:eqrates}
\vspace{-6pt}
\end{table}

\subsubsection{Evaluating the rates}
For a valid $(\alpha, \beta,\lambda)$ the achievable rates can be evaluated from Theorem \ref{thm:macfb_better} to be
\be \label{eq:rate_reg}
\begin{split}
R_1,R_2&<\min\{G,H\},\\
R_1+R_2 &<\frac{1}{2}\left(1+\frac{2P}{\sigma^2}(1+\overline{\alpha+\beta}+\lambda\beta) \right),
\end{split}
\ee
where
\ben
\begin{split}
G = &\frac{1}{2}\log\left(1+ \frac{\alpha P}{\sigma^2} + \frac{\beta P (1+\lambda)}{\alpha P + \sigma^2} \right),\\
H =& \frac{1}{2}\log(1+ \alpha\frac{P}{\sigma^2}) + \frac{1}{2}\log(1+ \frac{4 \:\overline{\alpha+\beta}\: {P}/\sigma^2}{2(\alpha+\beta+\beta\lambda)P/\sigma^2+1}) \\
& +\frac{1}{2}\log\left(1+ \frac{\beta (1+\lambda) P/\sigma^2}{(1+2\alpha P/\sigma^2) (1+\alpha P/\sigma^2)}\right).
\end{split}
\een
For different values of the signal-to-noise ratio ${P}/{\sigma^2}$, we  (numerically) compute the equal-rate point $(R^*,R^*)$ on the boundary of \eqref{eq:rate_reg}. For various values of  $P/\sigma^2$, Table \ref{tab:eqrates} compares $R^*$ with $R_{CL}$, the equal-rate point of the C-L region given by \eqref{eq:cl_eqpt}, and with the equal rate-point $R_{\text{FBcap}}$ on the boundary of the feedback capacity region \cite{Ozarow84}. We observe that our equal-rate points represent a significant improvement over the C-L region, and are close to the feedback capacity for large SNR.

\subsection{Comparison with Kramer's Generalization of the Cover-Leung Region} \label{subsec:kramer_comp}
In \cite[Section 5.3-5.4]{Kramer98}, a multi-letter generalization of the Cover-Leung region using was proposed.
This characterization was based on directed information, and is given below.
\begin{defi} For a triple of $M$-dimensional random vectors $(A^M,B^M, C^M)$ jointly distributed according
to $P_{A^M,B^M, C^M} = \prod_{i=1}^M P_{A_i,B_i,C_i|A^{i-1},B^{i-1},C^{i-1}}$, we define
\begin{gather}
I(A^M \to B^M) =\sum_{i=1}^M I(A^i;B_i|B^{i-1}), \\
I(A^M \to B^M || C^M) = \sum_{i=1}^M I(A^i;B_i|B^{i-1} \, C^i).
\end{gather}
The first quantity above is called the directed information from $A^M$ to $B^M$, and the second quantity is
the directed information from $A^M$ to $B^M$ causally conditioned on $C^M$. For any random variable $V$ jointly distributed
with these random vectors, the above definitions are extended in the natural way when we condition on $V$:
\begin{gather}
I(A^M \to B^M|V) =\sum_{i=1}^M I(A^i;B_i|B^{i-1} \ V), \\
I(A^M \to B^M || C^M | V) = \sum_{i=1}^M I(A^i;B_i|B^{i-1} \, C^i \, V).
\end{gather}
\end{defi}
\textbf{Fact $2$} (Generalized C-L region \cite{Kramer98}):
For any positive integer $M$, consider a joint distribution of the form
\ben
\begin{split}& P_{U^M X_1^M X_2^M Y^M}(u^M,x_1^M,x_2^M,y^M) =\\
&\prod_{i=1}^M P_U(u_i) \ P_{X_{1i}|U X_1^{i-1} Y^{i-1}}(x_{1i}|u_i \ x_1^{i-1} \ y^{i-1})\\
&\quad \cdot P_{X_{2i}|U X_2^{i-1} Y^{i-1}}(x_{2i}|u_i\ x_2^{i-1} \ y^{i-1}) \ P_{Y|X_1X_2}(y_i|x_{1i}\ x_{2i})
\end{split}
\een
where $P_{Y|X_1 X_2}$ is fixed by the channel, and the other distributions can be picked arbitrarily. Then the following rate pairs $(R_1,R_2)$ are achievable
over the MAC with noiseless feedback:
\be \label{eq:kramer_region}
\begin{split}
R_1 & \leq \frac{1}{M} I(X_1^M \to Y^M || X_2^M | U^M), \\
R_2 & \leq \frac{1}{M} I(X_2^M \to Y^M || X_1^M | U^M), \\
R_1 + R_2 & \leq \frac{1}{M} I(X_1^M X^M_2 \to Y^M). \\
\end{split}
\ee
We now compare the region of Theorem \ref{thm:macfb_better} with the generalized C-L region for $M=2$. This is a fair comparison  because in each of these regions, we have five distributions to pick: $P_U$, and two conditional distributions for each encoder. With $M=2$, the equal rate point on the boundary of \eqref{eq:kramer_region} was computed for a few examples in \cite{Kramer98}.  For the AWGN MAC with $P/\sigma^2=10$, the best equal rate pair was $R_1=R_2=1.2566$ bits, which is smaller than the rate $1.2709$ bits obtained using Theorem \ref{thm:macfb_better}  (see Table \ref{tab:eqrates}).

Consider the joint distribution of the generalized C-L scheme for $M=2$:
\ben
\begin{split}
& P_U(u_1) P_{X_{11}|U}(x_{11}|u_1) P_{X_{21}|U}(x_{21}|u_1) P_{Y|X_1X_2}(y_1|x_{11} x_{21}) \\
& P_U(u_2)  P_{X_{12}|U X_{11} Y_1}(x_{12}|u_2  x_{11}  y_1 ) P_{X_{22}|U X_{21} Y_1}(x_{22}|u_2 x_{21} y_1) \\
& P_{Y|X_1 X_2}(y_2|x_{12} x_{22}).
\end{split}
\een
The generalized C-L scheme  uses block-Markov superposition with $L$ blocks of transmission, each block being of length $N$. (Without loss of generality,
we will assume  that the block  length $N$ is even.) At the beginning of each block, to resolve the decoder's residual uncertainty, both encoders agree on the $U$ codeword $(u_1, \ldots, u_N)$, chosen i.i.d according to $P_U$. Each of the $2^{NR_1}$ codewords of encoder $1$ is generated according the following distribution:
\be
\begin{split}
&P_{X_{11}|U}(x_{11}|u_1)\: P_{X_{12}|U X_{11} Y_1}(x_{12}|u_2 \ x_{11} \ y_1 ) \:P_{X_{11}|U}(x_{13}|u_3)\\
&P_{X_{12}|U X_{11} Y_1}(x_{14}|u_3 \  x_{13} \ y_3) \ldots
\end{split}
\ee
In other words, the odd-numbered symbols of the block are chosen conditioned on just $U$ (like in the C-L scheme), while the even-numbered symbols are chosen conditioned on the preceding input symbol and the corresponding output. Equivalently, we can think of the block of length $N$ being divided into two sub-blocks of length $\frac{N}{2}$, where the first sub-block has symbols chosen i.i.d according to $P_{X_{11}|U}$, and the symbols of the second sub-block are chosen iid
according to $P_{X_{12}|U X_{11} Y }$, i.e., conditioned on the inputs and outputs of the first sub-block.

We can now establish an analogy between this coding scheme and that of Theorem \ref{thm:macfb_better}. In Theorem \ref{thm:macfb_better}, choose
$A= (\tilde{X}_1, \tilde Y)$ and $B= (\tilde{X}_2, \tilde{Y})$. (Recall that $\:\tilde{} \:$ is used to denote symbols of the previous block.)
It can be verified that the consistency condition \eqref{eq:q1q2cond} is trivially satisfied for this choice of $A$ and $B$.
With this choice, the encoder $1$ generates its inputs in each block according to $P_{X_1|U \tilde{X}_1 \tilde{Y}}$, and encoder $2$ generates its inputs according  to $P_{X_2|U \tilde{X}_2 \tilde{Y}}$. In particular, note that encoder $1$ chooses the channel inputs for  the \emph{entire} block conditioned on the channel outputs and its own inputs of the previous block.  In contrast, the generalized C-L scheme uses such a conditional input distribution only  for \emph{one half} of each block (the second sub-block). In the other half, the input symbols are conditionally independent given $U$. Since our coding scheme utilizes the correlation generated by feedback for the entire block, we expect it to yield higher rates. Of course, this comparison was made with the specific choice $A= (\tilde{X}_1, \tilde Y), \ B= (\tilde{X}_2, \tilde{Y})$. Other choices of $A$ and $B$ may yield higher rates in Theorem $1$ - the AWGN MAC in the previous subsection is such an example.

We emphasize that this is only a qualitative comparison of the two coding schemes, and we have not formally shown that generalized C-L region for $M=2$ is strictly contained in the rate region of Theorem \ref{thm:macfb_better} for the above choice of $A$ and $B$.

\subsection{Comparison with Bross-Lapidoth Region} \label{subsec:bl_comp}
Bross and Lapidoth (B-L) \cite{BrossLapi05}  established a rate region  that extends the Cover-Leung region.  The B-L scheme uses block-Markov superposition coding. Each block consists of two phases - a MAC phase and a two-way phase, and  is transmitted in $(1+\eta)N$ units of time. In the MAC phase of length $N$, the encoders send fresh information for the current block superimposed over  resolution information for the previous block. This part of the B-L scheme is identical to the Cover-Leung scheme. This is followed by the two-way phase of length $\eta N$ where the encoders communicate to exchange functions  $V_1$ and $V_2$ of the information available to each of them.

 In our coding scheme, $A$ and $B$ play a role similar to the functions $V_1$ and $V_2$  - they are generated based on the information available to the encoders at the end of the block. The key difference lies in how they are exchanged. In the B-L scheme, an \emph{extra} $\eta N$ time units is spent in each block to exchange $V_1,V_2$. Our scheme superimposes this information onto the next block; each block $l$ carries three layers of information - the base layer $U$ to resolve the decoder's list of block $(l-2)$, information exchange through $A$ and $B$ for the encoders to learn the messages of block $(l-1)$, and fresh messages corresponding to block $l$.

 Each block in our scheme has length $N$ as opposed to $(1+\eta)N $ in B-L, i.e., our scheme may be viewed as superimposing the two-way phase of the B-L scheme onto the MAC phase. In general, superposition is a more efficient way of exchanging correlated information than dedicating extra time for the exchange\footnote{For similar reasons, the Cover-Leung scheme outperforms the Gaarder-Wolf scheme for the binary erasure MAC \cite{GaardWolf75, CL81}.}; however, in order to obtain a single-letter rate region with superposition-based information exchange, we cannot choose $P_{AB}$ arbitrarily - it needs to satisfy the consistency condition \eqref{eq:q1q2cond}. Hence a direct comparison of our rate region with the Bross-Lapidoth region appears difficult.
Both the B-L region and our region are non-convex optimization problems, and there are no efficient ways to solve these. (In fact, the C-L region and the no-feedback MAC capacity region are non-convex optimization problems as well.)
In \cite{BrossLapi05}, the Poisson two-user MAC with feedback was considered as an example. It was shown that
computing the feedback capacity of the Poisson MAC is equivalent to computing the feedback capacity of the following binary MAC.
The  binary MAC, with inputs $(X_1,X_2)$ and output $Y$ is specified by
\begin{gather*}
P_{Y|X_1X_2}(1|01)=P_{Y|X_1X_2}(1|10)=q,\\
P_{Y|X_1X_2}(1|11)=2q, \; P_{Y|X_1X_2}(1|00)=0
\end{gather*}
where $0<q<0.5$. Note that if an encoder input is $0$ and the channel output is $1$, the other input is uniquely determined. In all other cases, one input, together with the output, does not determine the other input. Thus the condition for C-L optimality \cite{Willems82} is not satisfied.

It was shown in \cite{BrossLapi05} that feedback capacity region of the two-user Poisson MAC is the set of all rate pairs $\lim_{q \to 0} (\frac{R_1(q)}{q},\frac{R_2(q)}{q})$, where $(R_1(q), R_2(q))$ are achievable for the above binary MAC with feedback
achievable for the above binary channel with parameter $q$.
We shall compare the maximal equal rate points for this channel for small $q$.
The maximum symmetric sum rate in the C-L region is \cite{BrossLapi05}
\be
\frac{1}{q}(R_1+R_2)=0.4994 + o(1) \text{ nats}.
\ee
where $o(1)\to 0$ as $q\to 0$.
Our rate region from Theorem \ref{thm:macfb_better} yields the symmetric sum-rate
\be
\frac{1}{q}(R_1+R_2)=0.5132 + o(1) \text{ nats}.
\ee
The computation is found in Appendix \ref{app:compute}.  The  B-L symmetric sum rate reported in \cite{BrossLapi05} is
$\frac{1}{q}(R_1+R_2)=0.553 + o(1) \text{ nats}$, but there appears to be an error in the calculation, which we have communicated to the authors.

\section{Proof of Theorem \ref{thm:macfb_better}}\label{sec:proof}
\subsection{Preliminaries} \label{subsec:properties}
We shall use the notion of strong typicality as defined in \cite{CKbook}. Consider three finite sets $\mathcal{V}, \mathcal{Z}_1$ and
$\mathcal{Z}_2$, and an arbitrary distribution $P_{VZ_1Z_2}$ on them.
\begin{defi}
For any distribution $P_V$ on $\mathcal{V}$, a sequence $v^N \in \mathcal{V}^N$ is said to be $\epsilon$-typical with respect to $P_V$,
if
\[
\left| \frac{1}{N} \#(a|v^N) -P_V(a) \right| \leq \frac{\epsilon}{|\mathcal{V}|},
\]
for all $a \in \mathcal{V}$, and no $a \in \mathcal{V}$ with $P_V(a)=0$ occurs in $v^N$, where $\#(a|v^N)$ denotes the number of occurrences of $a$ in $v^N$.
Let $A^{(N)}_\epsilon(P_V)$ denote the set of all sequences that are $\epsilon$-typical with respect to $P_V$.
\end{defi}
The following are some of the properties of typical sequences that will be used in the proof.

\noindent \textbf{Property 0:} For all $\epsilon>0$, and for all sufficiently large $N$, we have $P_V^N[A^{(N)}_{\epsilon}(P_V)] > 1-\epsilon$.

\noindent \textbf{Property 1:} Let $v^N \in A^{(N)}_{\epsilon}(P_V)$ for some fixed $\epsilon>0$. If a random vector $Z_1^N$ is generated from the product distribution $\prod_{i=1}^N P_{Z_1|V}(\cdot|v_i)$, then  for all sufficiently large $N$, we have
$Pr[(v^N,Z_1^N) \not \in A^{(N)}_{\tilde{\epsilon}}(P_{VZ_1}) ]< \epsilon$, where $\tilde{\epsilon}=\epsilon(|\mathcal{V}|+|\mathcal{Z}_1|)$.

\noindent \textbf{Property 2:} Let $v^N \in A^{(N)}_{\epsilon}(P_V)$ for some fixed $\epsilon>0$. If a random vector $Z_1^N$ is generated from the product distribution $\prod_{i=1}^N P_{Z_1|V}(\cdot|v_i)$ and $Z_2^N$ is generated from the product distribution $\prod_{i=1}^N P_{Z_2|V}(\cdot|v_i)$, then for all sufficiently large
$N$, we have
\[\text{Pr}[(v^N,Z_1^N,Z_2^N) \in A^{(N)}_{\tilde{\epsilon}}(P_{VZ_1Z_2}) ]< \frac{2^{N\delta(\epsilon)}\ 2^{NH(Z_1Z_2|V)}}{2^{NH(Z_1|V)}  2^{NH(Z_2|V)}}\]
where
$\tilde{\epsilon}=\epsilon(|\mathcal{V}|+|\mathcal{Z}_1||\mathcal{Z}_2|)$,
and $\delta(\epsilon)$ is a continuous positive function of $\epsilon$ that goes to $0$ as $\epsilon \to 0$.

\subsection{Random Codebook generation} \label{subsec:randcodebook}
Fix a distribution $P_{UABX_1X_2Y}$ from $\mathcal{P}$ as in \eqref{eq:joint_dist1}, and a pair of conditional distribution $(Q_{A|\tilde{S},\tilde{X}_1},Q_{B|\tilde{S},\tilde{X}_2})$ from $\mathcal{Q}$. Fix  positive integers $N, M_1$ and $M_2$. $N$ is the block length, $M_1$ and $M_2$ denote the size of the message sets of the two transmitters in each block. Fix a positive integer $L$; $L$ is the number of blocks in encoding and decoding. Let $M_0[1]=M_0[2]=1$, and fix $(L-2)$ positive integers $M_0[l]$ for $l=3,\ldots,L$. Fix $\epsilon>0$, and let $\epsilon[l]=\epsilon(2|\mathcal{S}||\mathcal{X}_1| |\mathcal{X}_2|)^{l-1}$.

Recall that $S$ denotes the collection $(U,A,B,Y)$ and $\mathcal{S}$ denotes $\mathcal{U} \times \mathcal{A}
\times\mathcal{B} \times \mathcal{Y}$. For $l = 2,\ldots,L$, independently perform the following random
experiments.
\begin{itemize}
\item For every $(\mathbf{s},\mathbf{x}_1) \in \mathcal{S}^N \times \mathcal{X}_1^N$, generate one sequence
$\mathbf{A}_{[l,\mathbf{s},\mathbf{x}_1]}$ from $\prod_{n=1}^N Q_{A|\tilde{S},\tilde{X}_1}(\cdot|s_n,x_{1n})$.
\item Similarly, for every $(\mathbf{s},\mathbf{x}_2) \in \mathcal{S}^N \times \mathcal{X}_2^N$, generate one sequence
$\mathbf{B}_{[l,\mathbf{s},\mathbf{x}_2]}$
from $\prod_{n=1}^N Q_{B|\tilde{S},\tilde{X}_2}(\cdot|s_n,x_{2n})$.
\end{itemize}
For $l=1$, independently perform the following random experiment.
\begin{itemize}
\item Generate a pair  of sequences $(A^N_{[1,-,-]},B^N_{[1,-,-]})$ from the product distribution $P_{AB}^N$. The  dashes indicate that for the first block, $A^N$ and $B^N$ are generated directly using $P_{AB}$, unlike blocks $2,\ldots,L$ where they are generated using the $S,X_1,X_2$ sequences corresponding to the previous block.
\end{itemize}
\noindent For $l = 1,\ldots,L$, independently perform the following random
experiments.
\begin{itemize}
\item Independently choose $M_0[l]$ sequences $\mathbf{U}_{[l,m]}$, $m=1,2,\ldots,M_0[l]$, where each sequence is generated  from the product distribution $P_U^N$.
\item For each $(\mathbf{u},\mathbf{a}) \in \mathcal{U}^N \times \mathcal{A}^N$, independently generate $M_1$ sequences
$\mathbf{X}_{1[l,i,\mathbf{u},\mathbf{a}]}$, $i=1,2,\ldots,M_1$,  where each sequence is generated from $\prod_{n=1}^N P_{X_1|UA}(\cdot|u_n,a_n)$.
\item Similarly, for  each $(\mathbf{u},\mathbf{b}) \in \mathcal{U}^N \times \mathcal{B}^N$, independently generate $M_2$ sequences
$\mathbf{X}_{2[l,i,\mathbf{u},\mathbf{b}]}$,  $i=1,2,\ldots,M_2$, where each sequence is generated from $\prod_{n=1}^N P_{X_2|UB}(\cdot|u_n,b_n)$.
\end{itemize}

Upon receiving the channel output of block $l$, the  decoder decodes the message pair corresponding to block $(l-2)$, while the
encoders decode the messages of one another corresponding to block $(l-1)$. This is explained below.
\subsection{Encoding Operation}
Let $W_{1}[l]$ and $W_{2}[l]$ denote the transmitters' messages for block $l$. These are independent random variables uniformly distributed over
$\{1,2,\ldots,M_1\}$ and $\{1,2,\ldots,M_2\}$, respectively for $l=1,2,\ldots,(L-2)$. We set $W_1[0]=W_2[0]=W_1[L-1]=W_2[L-1]=W_1[L]=W_2[L]=1$.
For large, $L$, this will have a negligible effect on the rates.

For each block $l$, the encoder $1$ chooses a triple of sequences from $\mathcal{U}^N \times \mathcal{A}^N \times \mathcal{X}_1^N$, denoted by
$(\mathbf{U}_1[l],\mathbf{A}[l],\mathbf{X}_1[l])$, according to the encoding rule given below. Similarly encoder $2$ chooses a triple of sequences from $\mathcal{U}^N \times \mathcal{B}^N \times \mathcal{X}_2^N$ which is denoted  by $(\mathbf{U}_2[l],\mathbf{B}[l],\mathbf{X}_2[l])$. We will later see that with high probability $\mathbf{U}_1[l]=\mathbf{U}_2[l]$.

The MAC  output sequence in block $l$ is denoted by $\mathbf{Y}[l]$. Since output feedback is available at both the encoders, each encoder
maintains a copy of the decoder, so all three terminals are in synchrony.

{\small{
\begin{table*}[t]
\caption{Time-line of events for two successive blocks (each block of length $N$)}
 \label{tab:timeline}
\centering
\begin{tabular}{|c|cccccr|}
\hline
\tt{Time instant}  & $\ldots$ & $(l-1)N$ & $(l-1)N+1$  & $\ldots$ & $l N$ & $lN + 1$ \\
 &  &  block $(l-1)$ ends  & block $l$ begins &  & block $l$ ends  & block $(l+1)$ begins \\
\hline 
\tt{Encoder $1$ knows}  & $\ldots$ & $\mathbf{a}_{l-1}, W_{1(l-2)}, \mathbf{y}_{l-1}$ &   &  & $\mathbf{a}_{l}, W_{1(l-1)},\mathbf{y}_{l}$ & \\
\hline
%
\tt{Encoder $1$ decodes}  & $\ldots$ & $\mathbf{b}_{l-1} \to W_{2(l-2)}$ &   &  & $\mathbf{b}_{l} \to W_{2(l-1)}$ & \\
\hline 
\tt{Encoder $1$ produces}   & $\ldots$ & & $\mathbf{u}_l,\mathbf{a}_l \to \mathbf{x}_{1l}$  & &  & $\mathbf{u}_{l+1},\mathbf{a}_{l+1} \to \mathbf{x}_{1(l+1)}$\\
%
\hline \hline
\tt{Encoder $2$ knows}  & $\ldots$ & $\mathbf{b}_{l-1}, W_{2(l-2)}, \mathbf{y}_{l-1}$ &   &  & $\mathbf{b}_{l}, W_{2(l-1)}, \mathbf{y}_{l}$ & \\
\hline
\tt{Encoder $2$ decodes} & $\ldots$ & $\mathbf{a}_{l-1} \to W_{1(l-2)}$ &   &  & $\mathbf{a}_{l} \to W_{1(l-1)}$ & \\
\hline 
\tt{Encoder $2$ produces}  & $\ldots$ & & $\mathbf{u}_l,\mathbf{b}_l \to \mathbf{x}_{2l}$  & & & $\mathbf{u}_{l+1},\mathbf{b}_{l+1} \to \mathbf{x}_{2(l+1)}$ \\
\hline \hline
\tt{Decoder}   & &$\mathbf{u}_{l-1} \to$&   & & $\mathbf{u}_{l} \to$ & \\
\tt{decodes}   & & $W_{1(l-3)},W_{2(l-3)}$  & &  & $W_{1(l-2)},W_{2(l-2)}$& \\
\hline
\end{tabular}
\end{table*}
}}
\noindent \textbf{Block $1$:}
\begin{itemize}
\item Encoder 1 computes $\mathbf{U}_1[1]=\mathbf{U}_{[1,1]}$, $\mathbf{A}[1]=\mathbf{A}_{[1,-,-]}$, and $\mathbf{X}_1[1]=\mathbf{X}_{1[1,W_{1}[1],\mathbf{U}_1[1],\mathbf{A}[1]]}$. Then transmits $\mathbf{X}_1[1]$ as the channel input sequence.

\item Encoder 2 computes $\mathbf{U}_2[1]=\mathbf{U}_{[1,1]}$, $\mathbf{B}[1]=\mathbf{B}_{[1,-,-]}$, and
$\mathbf{X}_2[1]=\mathbf{X}_{2[1,W_{2}[1],\mathbf{U}_2[1],\mathbf{B}[1]]}$. Then transmits $\mathbf{X}_2[1]$ as the channel input sequence.

\item The MAC produces $\mathbf{Y}[1]$.

\item Encoder 1 sets $j[0]=1, \hat{\mathbf{B}}[1]=\mathbf{B}[1]$, and $\mathbf{S}_1[1]=(\mathbf{U}_1[1],\mathbf{A}[1],\hat{\mathbf{B}}[1],\mathbf{Y}[1])$.
For $l=1,\ldots,L$, $j[l]$ denotes encoder $1$'s estimate of $W_2[l]$, and $\hat{\mathbf{B}}[l]$ denotes its estimate of $\mathbf{B}[l]$.

\item Encoder 2 sets $i[0]=1, \hat{\mathbf{A}}[1]=\mathbf{A}[1]$, and $\mathbf{S}_2[1]=(\mathbf{U}_2[1],\hat{\mathbf{A}}[1],\mathbf{B}[1],\mathbf{Y}[1])$.\footnote{We see that $\mathbf{S}_1[1]=\mathbf{S}_2[1]$. In future blocks, this  will only hold with high probability.} For $l=1,\ldots,L$, $i[l]$ denotes encoder $1$'s estimate of $W_1[l]$, and $\hat{\mathbf{A}}[l]$ denotes its estimate of $\mathbf{A}[l]$.

\item Both encoders create  the list $\mathcal{L}[0]$ as the set containing the ordered pair $(1,1)$. $\mathcal{L}[l]$ denotes the list of highly likely message pairs corresponding to block $l$ at the decoder. The construction of this list for $l>1$ will be described in Section \ref{subsec:decoding}.
\end{itemize}
\begin{figure*}[!b]
\vspace*{2pt} \hrulefill
\normalsize 
\setcounter{mytempeqncnt}{\value{equation}}
\setcounter{equation}{28}
\be
\begin{split}
\mathcal{L}[1]= &\left\{(i,j):
(\mathbf{S}[1],\mathbf{X}_{1[1,i,\mathbf{U}_{[1,k[1]]},\bar{\mathbf{A}}[1]]},\mathbf{X}_{2[1,j,\mathbf{U}_{[1,k[1]]},\bar{\mathbf{B}}[1]]})
\ \ \mbox{is $\epsilon[l]$-typical and} \qquad \right. \\
& \qquad \left.
(\mathbf{U}_{[2,k[2]]}, \mathbf{Y}[2],\mathbf{A}_{[2,\mathbf{S}[1],\mathbf{X}_{1[1,i,\mathbf{U}_{[1,k[1]]},\bar{\mathbf{A}}[1]]}]}
\mathbf{B}_{[2,\mathbf{S}[1],\mathbf{X}_{2[1,j,\mathbf{U}_{[1,k[1]]},\bar{\mathbf{B}}[1]]}]})
\ \mbox{is $\epsilon[l]$-typical} \right\},
\end{split}
\label{eq:listL1}
\ee
\be
\begin{split}
\mathcal{L}[l-1]=&\left\{(i,j):
  (\mathbf{S}[l-1],\mathbf{X}_{1[l-1,i,\mathbf{U}_{[l-1,k[l-1]]},\bar{\mathbf{A}}[l-1]]},
\mathbf{X}_{2[l-1,j,\mathbf{U}_{[l-1,k[l-1]]},\bar{\mathbf{B}}[l-1]]})
  \ \ \mbox{is $\epsilon[l]$-typical and} \qquad \right. \\
&\quad \left.
(\mathbf{U}_{[l,k[l]]}, \mathbf{Y}[l],\mathbf{A}_{[l,\mathbf{S}[l-1],\mathbf{X}_{1[l-1,i,\mathbf{U}_{[l-1,k[l-1]]},\bar{\mathbf{A}}[l-1]]}]}
\mathbf{B}_{[l,\mathbf{S}[l-1],\mathbf{X}_{2[l-1,j,\mathbf{U}_{[l-1,k[l-1]]},\bar{\mathbf{B}}[l-1]]}]})
\ \mbox{is $\epsilon[l]$-typical} \right\}.
\end{split}
\label{eq:listLel}
\ee
\setcounter{equation}{\value{mytempeqncnt}} 
\addtocounter{equation}{2}
\end{figure*}
\noindent \textbf{Block $l$}, $l=2,\ldots,L$: The encoders perform the following sequence of operations.
\begin{itemize}
\item If the message pair $(W_{1}[l-2],j[l-2])$ is present in the list $\mathcal{L}[l-2]$, Encoder $1$ computes $k_1[l]$ as the index of this message pair in the list $\mathcal{L}[l-2]$. Otherwise, it sets $k_1[l]=1$. Encoder $1$ then computes $\mathbf{U}_1[l]=\mathbf{U}_{[l,k_1[l]]}$, $\mathbf{A}[l]=\mathbf{A}_{[l,\mathbf{S}_1[l-1],\mathbf{X}_1[l-1]]}$, and $\mathbf{X}_1[l]=\mathbf{X}_{1[l,W_{1}[l],\mathbf{U}_1[l],\mathbf{A}[l]]}$. It then transmits $\mathbf{X}_1[l]$ as the channel input sequence.

\item If the message pair $(i[l-2],W_2[l-2])$ is present in the list $\mathcal{L}[l-2]$, Encoder $2$ computes $k_2[l]$ as the index of this message pair in the list $\mathcal{L}[l-2]$. Otherwise, it sets $k_2[l]=1$. Encoder $2$ then computes $\mathbf{U}_2[l]=\mathbf{U}_{[l,k_2[l]]}$, $\mathbf{B}[l]=\mathbf{B}_{[l,\mathbf{S}_2[l-1],\mathbf{X}_2[l-1]]}$, and $\mathbf{X}_2[l]=\mathbf{X}_{2[l,W_{2}[l],\mathbf{U}_2[l],\mathbf{B}[l]]}$.
It then transmits $\mathbf{X}_1[l]$ as the channel input sequence.

\item The MAC produces $\mathbf{Y}[l]$.
\item
After receiving $\mathbf{Y}[l]$, Encoder $1$ wishes to decode
$W_{2}[l-1]$.  It attempts to find a unique index $j[l-1]$  such that the following two tuples:
{\small{
\ben
\begin{split}
&\mathbf{S}_1[l-1],\mathbf{X}_1[l-1],\mathbf{X}_{2[(l-1),j[l-1],\mathbf{U}_1[l-1],\hat{\mathbf{B}}[l-1]]} \text{ and}\\
& \mathbf{U}_1[l],\mathbf{A}[l],\mathbf{X}_{1}[l],\mathbf{Y}[l], \mathbf{B}_{[l,\mathbf{S}_1[l-1],\mathbf{X}_{2[l-1,j[l-1],\mathbf{U}_1[l-1],\hat{\mathbf{B}}[l-1]]}]}
\end{split}
\een
}}
are jointly $\epsilon[l]$-typical with respect to \eqref{eq:two_block_dist}. Note that encoder $1$ uses $(\mathbf{U}_1[l-1],\mathbf{U}_1[l])$ in place of
$(\mathbf{U}_2[l-1],\mathbf{U}_2[l])$ for this task. If there exists no such index or if
more than one such index is found, it sets $j[l-1]=1$. If
successful, it computes an estimate of $\mathbf{B}[l]$ using the
following equation:
\[\hat{\mathbf{B}}[l]=\mathbf{B}_{[l,\mathbf{S}_1[l-1],\mathbf{X}_{2[l-1,j[l-1],\mathbf{U}_1[l-1],\hat{\mathbf{B}}[l-1]]}]}.\]
It then computes  $\mathbf{S}_1[l]=(\mathbf{U}_1[l],\mathbf{A}[l], \hat{\mathbf{B}}[l],\mathbf{Y}[l])$.
\item
After receiving $\mathbf{Y}[l]$, Encoder $2$ wishes to decode $W_{1}[l-1]$.  It attempts to find a unique index $i[l-1]$  such that the following two tuples
{\small{
\ben
\begin{split}
& \mathbf{S}_2[l-1],\mathbf{X}_2[l-1],\mathbf{X}_{1[(l-1),i[l-1],\mathbf{U}_2[l-1],\hat{\mathbf{A}}[l-1]]}, \text{ and} \\
& \mathbf{U}_2[l],\mathbf{B}[l],\mathbf{X}_{2}[l],\mathbf{Y}[l], \mathbf{A}_{[l,\mathbf{S}_2[l-1],\mathbf{X}_{1[l-1,i[l-1],\mathbf{U}_2[l-1],\hat{\mathbf{A}} [l-1]]}]},
\end{split}
\een
}}
are jointly $\epsilon[l]$-typical with respect to \eqref{eq:two_block_dist}. Note that encoder $2$ uses $(\mathbf{U}_2[l-1],\mathbf{U}_2[l])$ in place of
$(\mathbf{U}_1[l-1],\mathbf{U}_1[l])$ for this task. If there exists no such index or if more than one such index is found, it sets $i[l-1]=1$. If
successful, it computes an estimate of  $\mathbf{A}[l]$ using the following equation:
\[\hat{\mathbf{A}}[l]=\mathbf{A}_{[l,\mathbf{S}_2[l-1],\mathbf{X}_{1[l-1,i[l-1],\mathbf{U}_2[l-1],\hat{\mathbf{A}}[l-1]]}]}.\]
It then computes
$\mathbf{S}_2[l]=(\mathbf{U}_2[l],\hat{\mathbf{A}}[l],\mathbf{B}[l],\mathbf{Y}[l])$.

\item Both encoders then execute the actions of the decoder corresponding to block $l$ (described in the next subsection). This step results in a list of message pairs $\mathcal{L}[l-1]$ of block $l-1$, defined in equation \eqref{eq:listLel} at the bottom of this page.
\end{itemize}
The time-line of events at the encoder for two successive blocks is shown in Table \ref{tab:timeline}.
\subsection{Decoding Operation} \label{subsec:decoding}
\noindent \textbf{Block $1$:}
\begin{itemize}
\item The decoder receives $\mathbf{Y}[1]$, and sets $k[1]=1$
\end{itemize}

\noindent \textbf{Block $2$:}
\begin{itemize}

\item Upon receiving $\mathbf{Y}[2]$,  the decoder sets $k[2]=1$. It then sets  $\bar{\mathbf{A}}[1]=\mathbf{A}[1]$, $\bar{\mathbf{B}}[1]=\mathbf{B}[1]$, and $\mathbf{S}[1]=(\mathbf{U}_{[1,k[1]]},\bar{\mathbf{A}}[1],\bar{\mathbf{B}}[1],\mathbf{Y}[1])$.

\item The decoder computes $\mathcal{L}[1]$, the list of message pairs defined by \eqref{eq:listL1} at the bottom of this page.
\end{itemize}

\noindent \textbf{Block $l, \: l=3,\ldots,L$:}
\begin{itemize}
\item Upon receiving $\mathbf{Y}[l]$, the decoder determines the unique
index $k[l] \in \{1,2,\ldots,M_0[l]\}$ such that \\
$(\mathbf{Y}[l],\mathbf{U}_{[l,k[l]]}, \mathbf{Y}[l-1],\mathbf{U}_{[l-1,k[l-1]]} )$ is $\epsilon[l]$-typical. If no
such index exists or more than one such index exists, then the
decoder declares error.
If successful in the above operation, the decoder computes the $k[l]$th
pair in the list $\mathcal{L}[l-2]$, and declares it as the
reconstruction $(\hat{W}_{1}[l-2],\hat{W}_2[l-2])$ of the message pair.

\item The decoder computes an estimate of $\mathbf{A}[l-1]$ using the equation
\[
\bar{\mathbf{A}}[l-1]=\mathbf{A}_{[l-1,\mathbf{S}[l-2],\mathbf{X}_{1[l-2,\hat{W}_{1}[l-2],\mathbf{U}_{[l-2,k[l-2]]},\bar{\mathbf{A}}[l-2]]}]}.
\]
Similarly, the decoder computes an estimate of $\mathbf{B}[l-1]$ using
the equation
\[
\bar{\mathbf{B}}[l-1]=\mathbf{B}_{[l-1,\mathbf{S}[l-2],\mathbf{X}_{2[l-2,\hat{W}_{2}[l-2],\mathbf{U}_{[l-2,k[l-2]]},\bar{\mathbf{B}}[l-2]]}]}.
\]
The decoder then computes
\[
\mathbf{S}[l-1]=(\mathbf{U}_{[l-1,k[l-1]]}, \bar{\mathbf{A}}[l-1],
\bar{\mathbf{B}}[l-1],\mathbf{Y}[l-1]).
\]

\item The decoder then computes $\mathcal{L}[l-1]$, the list of message pairs defined by \eqref{eq:listLel} at the bottom of the previous page.
\end{itemize}

\subsection{Error Analysis}
For block $l \in \{1,2,\ldots,L\}$, if $\mathbf{U}_1[l]=\mathbf{U}_2[l]$, then let $\mathbf{U}[l]=\mathbf{U}_1[l]$; otherwise, let $\mathbf{U}[l]$ be a fixed deterministic sequence that does not depend on $l$.

\subsubsection*{Block 1}
Let $E[1]^c$ be the event that $(\mathbf{U}[1],\mathbf{A}[1],\mathbf{B}[1],\mathbf{X}_1[1],\mathbf{X}_2[1],\mathbf{Y}[1])$
is $\epsilon[1]$-typical with respect to $P_{UABX_1X_2Y}$. By Property $0$ in Section \ref{subsec:properties}, we have $\text{Pr}[E[1]] \leq \epsilon$ for all
sufficiently large $N$.

\subsubsection*{Block 2}
\begin{itemize}
\item[-] Let $E_1[2]$ be the event that Encoder $1$ fails to decode $W_{2}[1]$ upon receiving $\mathbf{Y}[2]$.
\item[-] Let $E_2[2]$ be the event that Encoder $2$ fails to decode $W_{1}[1]$ upon receiving $\mathbf{Y}[2]$.
\item[-] Let $E_3[2]$ be the event that at the decoder \[ |\mathcal{L}[1]|>2^{N(I(U;Y|\tilde U \tilde Y)-2\delta_1(\epsilon[2]))}. \] Here $\delta_1(\cdot)$  is a continuous positive function that tends to $0$ as its argument tends to $0$,  similar to the one used in Property $2$ of typical sequences.
\end{itemize}
    The error event $E[2]$ in Block $2$ is given by \[E[2]=E_1[2] \cup E_2[2] \cup E_3[2].\]

Conditioned on the event $E[1]^c$, the conditional probability that the tuples $(\mathbf{U}[1],\mathbf{A}[1],\mathbf{B}[1],\mathbf{X}_1[1],\mathbf{X}_2[1],\mathbf{Y}[1])$ and $(\mathbf{U}[2],\mathbf{A}[2],\mathbf{B}[2],$ $\mathbf{X}_1[2],\mathbf{X}_2[2],\mathbf{Y}[2])$ are not jointly $\epsilon[2]$-typical with respect to \eqref{eq:two_block_dist} is smaller than $\epsilon$ for all sufficiently large $N$ (by Property $1$). Using this and Property $2$ of typical sequences,
we have the following upper bound on $\text{Pr}[E_1[2]|E[1]^c]$:
\begin{equation}
\begin{split}
& \text{Pr}[E_1[2]|E[1]^c] \leq \epsilon + \sum_{j=1}^{M_2} \frac{2^{N \delta_1(\epsilon[2])} \ 2^{NH(\tilde{X}_2B| \tilde{S} \tilde{X}_1UAX_1Y)}}
{2^{NH(\tilde{X}_2|\tilde{U}\tilde{B})} \ 2^{NH(B|\tilde{S}\tilde{X}_2)}} \\
&\overset{(a)}{=} \epsilon + \sum_{j=1}^{M_2} 2^{N \delta_1(\epsilon[2])}
  2^{-N \left(I(\tilde{X}_2;\tilde{Y}|\tilde{U} \tilde{A} \tilde{B} \tilde{X}_1) + I(\tilde{X}_2B;Y|\tilde{S} \tilde{X}_1UAX_1)\right)}  \\
&\overset{(b)}{=} \epsilon + \sum_{j=1}^{M_2} 2^{N \delta_1(\epsilon[2])}
  2^{-N(I(X_2;Y|UA BX_1) + I(\tilde{X}_2 B;Y|\tilde{S} \tilde{X}_1UAX_1))}  \\
&\overset{(c)}{=} \epsilon + \sum_{j=1}^{M_2} 2^{N \delta_1(\epsilon[2])} 2^{-NI(X_2;Y|U A X_1  \tilde{X}_1 \tilde{S})} \overset{(d)}{\leq} 2 \epsilon.
\end{split}
\end{equation}
In the above, ($a$) can be obtained using the chain rule of mutual information along with the following Markov chains:
\begin{equation}
\begin{split}
 B \tilde{X}_2 - \tilde{S} \tilde{X}_1  - UA, \quad & \tilde{S} \tilde{X}_1 \tilde{X}_2 - UA - X_1,\\
 A \tilde{X}_1 - \tilde{S} \tilde{X}_2  - UB, \quad & \tilde{S} \tilde{X}_1 \tilde{X}_2 - UB - X_2, \\
  \tilde{S} \tilde{X}_1 \tilde{X}_2 -  UAB&  -  X_1 X_2 - Y.
\end{split}
\label{eq:markovchain}
\end{equation}
Indeed,
\begin{equation*}
\begin{split}
&H(\tilde{X}_2|\tilde{U}\tilde{B}) + H(B|\tilde{S}\tilde{X}_2) - H(\tilde{X}_2B| \tilde{S} \tilde{X}_1UAX_1Y)\\
&=I(\tilde{X}_2;\tilde{A}\tilde{X}_1\tilde{Y} UAX_1 Y|\tilde{U} \tilde{B}) +  I(B; \tilde{X}_1UAX_1Y|  \tilde{S} \tilde{X}_2)\\
&=I(\tilde{X}_2;\tilde{A}\tilde{X}_1\tilde{Y}|\tilde{U} \tilde{B}) + I(\tilde{X}_2; UAX_1Y| \tilde{S} \tilde{X}_1) \\
& \qquad +  I(B; UAX_1Y|\tilde{S} \tilde{X}_1 \tilde{X}_2)\\
&= I(\tilde{X}_2;\tilde{Y}|\tilde{U} \tilde{A} \tilde{B} \tilde{X}_1) + I(\tilde{X}_2 B;Y|\tilde{S} \tilde{X}_1UAX_1).
\end{split}
\end{equation*}
($b$) follows from the fact that
$(\tilde{S},\tilde{X}_1,\tilde{X}_2)$ has the same distribution as
$(S,X_1,X_2)$, and ($c$) can be obtained as follows using \eqref{eq:markovchain}:
\ben
\begin{split}
& I(X_2;Y|UA BX_1) + I(\tilde{X}_2 B;Y|\tilde{S} \tilde{X}_1UAX_1)\\
& = I(X_2;Y|UA BX_1 \tilde S \tilde X_1 \tilde X_2) + I(\tilde{X}_2 B;Y|\tilde{S} \tilde{X}_1UAX_1) \\
& = I(X_2 \tilde X_2 B; Y|\tilde S \tilde X_1 U A X_1) = I(X_2;Y|U A X_1 \tilde X_1 \tilde S).
\end{split}
\een
($d$) holds for all sufficiently large $N$ if
\begin{equation}
\frac{1}{N} \log M_2 < I(X_2;Y|\tilde{S} \tilde{X}_1UAX_1)-4\delta_1(\epsilon[2]).
\label{eq:cond1}
\end{equation}
Similarly
$\text{Pr}[E_2[2]|E[1]^c] \leq 2 \epsilon$ for all sufficiently large $N$ if
\begin{equation}
\frac{1}{N} \log M_1 < I(X_1;Y|\tilde{S}\tilde{X}_2 UBX_2)  - 4\delta_1(\epsilon[2]).
\label{eq:cond2}
\end{equation}

To bound $\text{Pr}[E_3[2]|E[1]^c]$, start by defining $\Psi_{k,l}=1$ if $(k,l) \in
\mathcal{L}[1]$ and equal to $0$ otherwise.  Then
\be \label{eq:list_size}
\begin{split}
E(|\mathcal{L}[1]|)= & E \Psi_{W_1[1],W_2[1]} + \sum_{i \neq W_1[1]} E\Psi_{i,W_2[1]} \\
& +\sum_{j \neq W_2[1]} E\Psi_{W_1[1],j}+\sum_{i \neq W_1[1],  j \neq W_2[1]} \hspace{-8pt} E\Psi_{i,j}.
\end{split}
\ee
For $j \neq
W_2[1]$, using Property $2$ of typical sequences we have
\begin{equation}
\begin{split}
\label{eq:psiw1}
& E\Psi_{W_1[1],j} \leq \frac{2^{N \delta_1(\epsilon[2])} \ 2^{NH(\tilde{X}_2B| \tilde{S} \tilde{X}_1UAY)}}
  {2^{NH(\tilde{X}_2|\tilde{U}\tilde{B})} \ 2^{NH(B|\tilde{S}\tilde{X}_2)}}   \\
&\stackrel{(a)}{=} 2^{N \delta_1(\epsilon[2])} \ 2^{-NI(\tilde{X}_2;\tilde{Y}|\tilde{U} \tilde{A} \tilde{B} \tilde{X}_1)}
\ 2^{-NI(B;Y|\tilde{S} \tilde{X}_1UA)}
\end{split}
\end{equation}
where $(a)$ is obtained by using the chain rule of mutual information and the Markov chains in \eqref{eq:markovchain} as follows.
{\small{
\begin{equation*}
\begin{split}
&H(\tilde{X}_2|\tilde{U}\tilde{B})+ H(B|\tilde{S}\tilde{X}_2) - H(\tilde{X}_2B| \tilde{S} \tilde{X}_1UAY) \\
&=I(\tilde X_2; \tilde A \tilde X_1 \tilde Y U A Y|\tilde U \tilde B) + I(B ;\tilde X_1 UAY| \tilde S \tilde X_2) \\
&={I(\tilde X_2; \tilde Y| \tilde U \tilde A \tilde B \tilde X_1) + I(\tilde X_2 ; Y| \tilde S \tilde X_1 U A)
+ I(B;Y|\tilde S \tilde X_1 \tilde X_2 U A)}\\
&= I(\tilde X_2; \tilde Y| \tilde U \tilde A \tilde B \tilde X_1) +  I(B \tilde X_2 ; Y| \tilde S \tilde X_1 U A) \\
& = I(\tilde X_2; \tilde Y| \tilde U \tilde A \tilde B \tilde X_1) +  I(B  ; Y| \tilde S \tilde X_1 U A).
\end{split}
\end{equation*}
}}
Using the fact that $(\tilde{S},\tilde{X}_1,\tilde{X}_2)$ has the
same distribution as $(S,X_1,X_2)$, \eqref{eq:psiw1} becomes
\be \label{eq:l1i}
\begin{split}
&\frac{1}{N} \log (\sum_{j \neq W_2[1]} \hspace{-4pt} E\Psi_{W_1[1],j} ) \leq  \frac{\log M_2}{N} -I(X_2;Y|U A B X_1) \\
&\qquad - I(B;Y|\tilde{S} \tilde{X}_1UA)  +\delta_1(\epsilon[2]).
\end{split}
\ee
Similarly,
\be \label{eq:l1j}
\begin{split}
& \frac{1}{N} \log(\sum_{i \neq W_1[1]} \hspace{-4pt}E\Psi_{i,W_2[1]}) \leq  \frac{\log M_1}{N}-I(X_1;Y|UABX_2) \\
& \qquad -I(A;Y|\tilde{S}\tilde{X}_2 UB) +\delta_1(\epsilon[2]).
\end{split}
\ee
Using Property $2$ of typical sequences, we have for $i \neq W_1[1]$ and $j \neq W_2[1]$,
\begin{equation}
\begin{split}
E\Psi_{i,j} &\leq \frac{2^{N\delta_1(\epsilon[2])} \ 2^{NH(\tilde{X}_1 \tilde{X}_2 AB| \tilde{S} UY)}}
{2^{NH(\tilde{X}_1|\tilde{U} \tilde{A})} \ 2^{NH(\tilde{X}_2|\tilde{U} \tilde{B})}
\ 2^{NH(A|\tilde{S} \tilde{X}_1)} \ 2^{NH(B|\tilde{S} \tilde{X}_2)}
} \\
&\stackrel{(a)}{=} 2^{N\delta_1(\epsilon[2])} \ 2^{-NI(\tilde{X}_1 \tilde{X}_2 ;\tilde{Y}| \tilde U \tilde A \tilde B)} 2^{-NI(AB;Y|U\tilde{S})}
\end{split}
\end{equation}
where $(a)$ is obtained by using the chain rule of mutual information and the Markov chains in \eqref{eq:markovchain} following steps similar to those for
\eqref{eq:psiw1}. Hence
\be \label{eq:l1ij}
\begin{split}
& \frac{1}{N} \log (\sum_{i \neq W_1[1], j \neq W_2[1]} \hspace{-6pt} E\Psi_{i,j}) \leq
\frac{1}{N} \log M_1+ \frac{1}{N} \log M_2 \\
& \qquad -I(X_1X_2;Y|UAB) -I(AB;Y|U \tilde{S}) +\delta_1(\epsilon[2]).
\end{split}
\ee

Using \eqref{eq:l1i},\eqref{eq:l1j} and \eqref{eq:l1ij}, \eqref{eq:list_size} can be written as
\be \label{eq:elb_bound}
\begin{split}
E|\mathcal{L}_{1}| \leq &  1+ M_1 2^{ -N(I(X_1;Y|UABX_2)  +I(A;Y|\tilde{S}\tilde{X}_2 UB) -\delta_1(\epsilon[2]))}\\
& + M_2 2^{-N(I(X_2;Y|UABX_1)  +I(B;Y|\tilde{S} \tilde{X}_1UA)   -\delta_1(\epsilon[2]))}  \\
&+M_1 M_2  2^{-N(I(X_1X_2;Y|UAB)  +I(AB;Y|U \tilde{S}) -\delta_1(\epsilon[2]))}.
\end{split}
\ee
Using \eqref{eq:elb_bound} in the Markov inequality, one can show that for all sufficiently large $N$,
\[ P\left(|\mathcal{L}[1]|<2^{N(\max\{T_1,T_2,T_3\} +2\delta_1(\epsilon[2]))}\right) > 1-\epsilon\]
where
{\small{
\be \label{eq:T1T2T3}
\begin{split}
T_1 \triangleq& \frac{\log M_1}{N} + \frac{\log M_2}{N} -I(X_1X_2;Y|ABU)-I(AB;Y|U\tilde{S}),\\
T_2 \triangleq& \frac{\log M_1}{N} -I(X_1;Y|X_2ABU)-I(A;Y|UB\tilde{S}\tilde{X}_2),\\
T_3 \triangleq& \frac{\log M_2}{N} -I(X_2;Y|X_1ABU)-I(B;Y|UA \tilde{S}\tilde{X}_1).
\end{split}
\ee }}
Hence $\text{Pr}[E_3[2]|E[1]^c] <2\epsilon$ if
 \be\label{eq:r1r2bound1}
\begin{split}
 \max\{T_1,T_2,T_3\} \leq  I(U;Y|\tilde U \tilde Y)-4\delta_1(\epsilon[2])
\end{split}
\ee
Hence $\text{Pr}[E[2]|E[1]^c] <  6\epsilon$ if \eqref{eq:cond1}, \eqref{eq:cond2} and \eqref{eq:r1r2bound1} are satisfied.
\subsubsection*{Block $l:3,\ldots,L$}
\begin{itemize}
\item[-] Let $E_1[l]$ be the event that after receiving $\mathbf{Y}[l]$, Encoder 1 fails to decode $W_{2}[l-1]$.
\item[-] Let $E_2[l]$ be the event that after receiving $\mathbf{Y}[l]$,  Encoder 2 fails to decode $W_{1}[l-1]$.
\item[-] Let $E_3[l]$ be the event that at the decoder $|\mathcal{L}[l-1]|> 2^{n(I(U;Y| \tilde U \tilde Y)-2\delta_1(\epsilon[l]))}$.
\item[-] Let $E_4[l]$ be the event that the decoder fails to correctly decode $\mathbf{U}[l]$.
\end{itemize}
The error event $E[l]$ in Block $l$ is given by \[ E[l]=E_1[l] \cup E_2[l] \cup E_3[l] \cup E_4[l]. \]

Using arguments similar to those used in Block 2, it can be shown that $\text{Pr}[E_i[l]|E[l-1]^c] < 2\epsilon$ for $i=1,2,3$ for all sufficiently large
$N$, if the conditions given by \eqref{eq:cond1}, \eqref{eq:cond2}, and \eqref{eq:r1r2bound1} are satisfied with $\epsilon[2]$ replaced by $\epsilon[l]$.
Moreover, using standard arguments one can also show that $\text{Pr}[E_4[l]|E[l-1]^c]<2\epsilon$ for all sufficiently large $N$ if
\be \label{eq:M0l}
\begin{split}
\frac{1}{N} \log M_0[l]&=I(U;\tilde U \tilde Y Y)-2\delta_1(\epsilon[l])\\
&= I(U;Y|\tilde U \tilde Y)-2\delta_1(\epsilon[l]).
\end{split}
\ee
Hence $\text{Pr}[E[l]|E[l-1]^c] < 8\epsilon$ for all sufficiently large $N$.
\subsubsection*{Overall Decoding Error Probability}
The above arguments imply that we can make the probability of decoding error over $L$  blocks  satisfy
\[
\text{Pr}[E]=\text{Pr}\left[ \bigcup_{l=1}^L E[l] \right] \leq 8L \epsilon
\]
if $M_0[l]$  is chosen according  \eqref{eq:M0l} for $l=3,\ldots,L$, and $M_1, M_2$ satisfy the following conditions:
\begin{align*}
&\frac{1}{N} \log M_2 \leq I(X_2;Y|\tilde{S} \tilde{X}_1UAX_1)-\theta\\
&\frac{1}{N} \log M_1 \leq I(X_1;Y|\tilde{S}\tilde{X}_2 UBX_2) -\theta
\end{align*}
\begin{equation*}
\begin{split}
&\frac{1}{N} \log M_1 + \frac{1}{N} \log M_2 \leq I(X_1X_2;Y|ABU)\\
& \qquad  +I(AB;Y|U\tilde{S}) + I(U;Y|\tilde U \tilde Y)-\theta\\
&\frac{1}{N} \log M_1 \leq I(X_1;Y|X_2ABU)+I(A;Y|UB\tilde{S}\tilde{X}_2)\\
& \qquad +I(U;Y|\tilde U \tilde Y)-\theta\\
&\frac{1}{N} \log M_2 \leq I(X_2;Y|X_1ABU)+I(B;Y|UA\tilde{S}\tilde{X}_1)\\
& \qquad +I(U;Y|\tilde Y \tilde Y)-\theta
\end{split}
\end{equation*}
where $\theta=\sum_{l=1}^{L} 4\delta_1(\epsilon[l])$. This implies that the following rate region is achievable.
{\small{
\begin{equation}
\begin{split}
&R_1  \leq I(X_1;Y|UABX_2)+I(A;Y|UB \tilde{S} \tilde{X}_2) +I(U;Y|\tilde U \tilde Y), \\
&R_2  \leq I(X_2;Y|UABX_1)+I(B;Y|UA \tilde{S} \tilde{X}_1) +I(U;Y|\tilde U \tilde Y), \\
&R_1  \leq I(X_1;Y|UBX_2 \tilde{S} \tilde{X}_2), \\
&R_2  \leq I(X_2;Y|UAX_1 \tilde{S} \tilde{X}_1), \\
&R_1+R_2  \leq I(X_1X_2;Y|ABU)+I(AB;Y|U \tilde{S})+I(U;Y|\tilde U \tilde Y).
\end{split}
\label{eq:achievable_region}
\end{equation}
}}

Next we show that the above rate region is equivalent to that given in Theorem \ref{thm:macfb_better}. Using the Markov chains in \eqref{eq:markovchain}, we get
\begin{equation}
\begin{split}
& I(X_1 X_2;Y|ABU) + I(AB;Y|U \tilde{S})\\
&=I(X_1 X_2;Y|ABU\tilde{S}) + I(AB;Y|U \tilde{S}) \\
&=I(AB X_1 X_2;Y|U\tilde{S})=I(X_1 X_2;Y|U\tilde{S}).
\end{split}
\label{eq:final0v}
\end{equation}
Moreover,
\begin{equation}
\begin{split}
&I(X_1;Y|UABX_2)+I(A;Y|UB \tilde{S} \tilde{X}_2)\\
&= I(X_1;Y|UABX_2 \tilde{S} \tilde{X}_2)+I(A;Y|UB \tilde{S} \tilde{X}_2) \\
&=I(X_1;YX_2|UAB\tilde{S} \tilde{X}_2)+I(A;Y|UB \tilde{S} \tilde{X}_2) \\
&= I(X_1;YX_2|UAB\tilde{S} \tilde{X}_2) +I(A;YX_2|UB \tilde{S} \tilde{X}_2) \\
& \qquad -I(A;X_2|UBY \tilde{S} \tilde{X}_2) \\
&= I(AX_1;YX_2|UB\tilde{S} \tilde{X}_2)-I(A;X_2|UBY\tilde{S} \tilde{X}_2) \\
&= I(AX_1;Y|UBX_2 \tilde{S} \tilde{X}_2) -I(A;X_2|UBY\tilde{S}\tilde{X}_2) \\
&= I(X_1;Y|UBX_2\tilde{S}\tilde{X}_2) -I(A;X_2|UBY\tilde{S}\tilde{X}_2).
\end{split}
\label{eq:final1v}
\end{equation}
Similarly,
\begin{equation}
\begin{split}
& I(X_2;Y|UABX_1) + I(B;Y|UA \tilde{S} \tilde{X}_1) \\
&=I(X_2;Y|UAX_1\tilde{S}\tilde{X}_1)-I(B;X_1|UAY\tilde{S}\tilde{X}_1).
\end{split}
\label{eq:final2v}
\end{equation}
\eqref{eq:final0v}, \eqref{eq:final1v} and \eqref{eq:final2v} imply the desired result.

\section{Extension of Coding Scheme}\label{sec:three_stage}
 We can extend the coding scheme by thinning the fully-connected graph to the perfectly correlated graph over \emph{three} blocks, i.e., going through two intermediate steps with progressively thinner graphs in each step. This yields a potentially larger rate region, as described below. Let the rate pair $(R_1,R_2)$ lie outside the region of Theorem \ref{thm:macfb_better}. Consider the transmission of message pair $(W_{1l}, W_{2l})$  through $(\mathbf{X}_{1l},\mathbf{X}_{2l})$ in block $l$. 

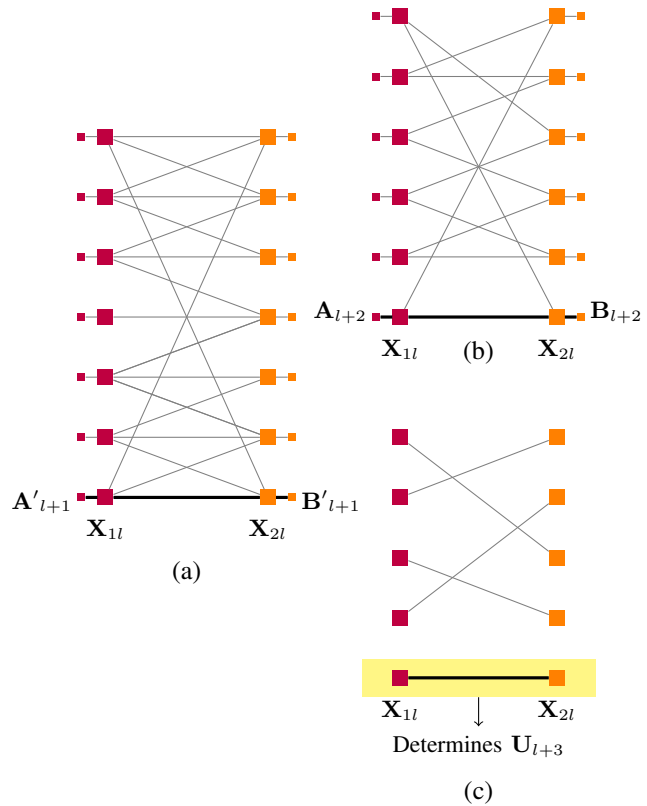
\begin{figure}
\centering
\begin{tikzpicture}[scale=0.8]
\foreach \x in {9,8,...,3}
{
\node[leftvertex] (L\x) at (-0.1,\x)  {};
\node[Sleftvertex] (LA\x) at (-0.5,\x) {};
\node[rightvertex] (R\x) at (2.6,\x)  {};
\node[Srightvertex] (RB\x) at (3.0,\x) {};
\draw [-, thin, gray] (L\x) to (R\x); \draw [-, thin, gray] (L\x) to (LA\x); \draw [-, thin, gray] (R\x) to (RB\x);
}
\draw [-, very thick] (L3) to node[below=20pt] {(a)} (R3);
\draw [-, very thick] (L3) to (LA3);
\draw [-, very thick] (R3) to (RB3);
\draw [-, thin, gray] (L3) to (R4); \draw [-, thin, gray] (L3) to (R9);
\draw [-, thin, gray] (L4) to (R5); \draw [-, thin, gray] (L4) to (R3);
\draw [-, thin, gray] (L5) to (R6); \draw [-, thin, gray] (L5) to (R4);
\draw [-, thin, gray] (L5) to (R6); \draw [-, thin, gray] (L5) to (R4);
\draw [-, thin, gray] (L7) to (R8); \draw [-, thin, gray] (L7) to (R6);
\draw [-, thin, gray] (L8) to (R9); \draw [-, thin, gray] (L8) to (R7);
\draw [-, thin, gray] (L9) to (R3); \draw [-, thin, gray] (L9) to (R8);
\node [below=2pt] at (L3.south) {\small{$\mathbf{X}_{1l}$}};
\node [below=2pt] at (R3.south) {\small{$\mathbf{X}_{2l}$}};
\node [left] at (LA3.south) {\small{$\mathbf{A'}_{l+1}$}};
\node [right] at (RB3.south) {\small{$\mathbf{B'}_{l+1}$}};
\foreach \x in {11,...,6}
{
\node[leftvertex] (L\x) at (4.8,\x)  {};
\node[Sleftvertex] (LA\x) at (4.4,\x) {};
\node[rightvertex] (R\x) at (7.4,\x)  {};
\node[Srightvertex] (RB\x) at (7.8,\x) {};
\draw [-, thin, gray] (L\x) to (LA\x); \draw [-, thin, gray] (R\x) to (RB\x);
}
\draw [-, very thick] (L6) to node[below=5pt] {(b)} (R6);
\draw [-, very thick] (L6) to (LA6); \draw [-, very thick] (R6) to (RB6);
\draw [-, thin, gray] (L7) to (R8); \draw [-, thin, gray] (L7) to (R7);
\draw [-, thin, gray] (L8) to (R9); \draw [-, thin, gray] (L8) to (R7);
\draw [-, thin, gray] (L9) to (R10); \draw [-, thin, gray] (L9) to (R8);
\draw [-, thin, gray] (L10) to (R11); \draw [-, thin, gray] (L10) to (R10);
\draw [-, thin, gray] (L11) to (R6); \draw [-, thin, gray] (L6) to (R11);
\draw [-, thin, gray] (L11) to (R9);
\node [below=2pt] at (L6.south) {\small{$\mathbf{X}_{1l}$}};
\node [below=2pt] at (R6.south) {\small{$\mathbf{X}_{2l}$}};
\node [left] at (LA6.north) {\small{$\mathbf{A}_{l+2}$}};
\node [right] at (RB6.north) {\small{$\mathbf{B}_{l+2}$}};
\foreach \x in {4,...,0}
{
\node[leftvertex] (L\x) at (4.8,\x)  {};
\node[rightvertex] (R\x) at (7.4,\x)  {};
}
\draw [-, thin, gray] (L4) to (R2);
\draw [-, thin, gray] (L3) to (R4);\draw [-, thin, gray] (L2) to (R1);
\draw [-, thin, gray] (L1) to (R3); \draw [-, thin, gray] (L0) to (R0);
\draw [-, very thick] (L0) to node[below=35pt] {(c)} (R0);
\node [below=2pt] at (L0.south) {\small{$\mathbf{X}_{1l}$}};
\node [below=2pt] at (R0.south) {\small{$\mathbf{X}_{2l}$}};
\begin{pgfonlayer}{background}
\node [fill=yellow!60, fit=(L0) (R0)] (BG) {};
\end{pgfonlayer}
\node [below=11pt] (txt) at (BG.south) {\small{Determines $\mathbf{U}_{l+3}$}};
\draw [->] (BG) -- (txt);
\end{tikzpicture}
\caption{Decoder's message graph for message pair $(W_{1l}, W_{2l})$:  a) After receiving $\mathbf{Y}_{l}$
b) After receiving $\mathbf{Y}_{l+1}$ c)After receiving $\mathbf{Y}_{l+2}$}
\label{fig:four_stage}
\end{figure}

\begin{itemize}
\item  At the end of block $l$, the effective message graph of the decoder given $\mathbf{Y}_b$ is shown in Figure \ref{fig:four_stage}(a). This is a correlated message graph. For each sequence $\mathbf{X}_{1}$, choose one sequence $\mathbf{A}'$, conditioned on the information at encoder $1$. Similarly, choose one sequence $\mathbf{B}'$  for each $\mathbf{X}_{2}$, based on the information at encoder $2$. The $\mathbf{A}'$ and $\mathbf{B}'$ sequences corresponding to $\mathbf{X}_{1l}$ and $\mathbf{X}_{2l}$ are set to $\mathbf{A}'_{l+1}$ and $\mathbf{B}'_{l+1}$, respectively. Note that $\mathbf{A}'$ and $\mathbf{B}'$ here are similar to $\mathbf{A}$ and $\mathbf{B}$ of the original coding scheme.

\item At the end of block $(l+1)$, both encoders and the decoder receive $\mathbf{Y}_{l+1}$. The degree of each left vertex in the graph of Figure \ref{fig:four_stage}(a) is too large for encoder $2$ to decode $\mathbf{A}'_{l+1}$ from $\mathbf{Y}_{l+1}$. Similarly, encoder $1$ cannot decode $\mathbf{B}'_{l+1}$ from $\mathbf{Y}_{l+1}$. So we have the correlated message graph of Figure \ref{fig:four_stage}(b)- this graph is a subgraph of the graph in Figure \ref{fig:four_stage}(a). An edge in graph \ref{fig:four_stage}(a) is present in graph \ref{fig:four_stage}(b) if and only if the corresponding $(\mathbf{A}'_{l+1},\mathbf{B}'_{l+1})$ pair is jointly typical with $\mathbf{Y}_{l+1}$. At the end of block $(l+1)$, though the encoders do not know the edge $(W_{1l},W_{2l})$, observe that we have \emph{thinned} the message graph, i.e., the degree of each vertex in graph \ref{fig:four_stage}(b) is strictly smaller than its degree in graph \ref{fig:four_stage}(a).

\item Each left vertex in graph \ref{fig:four_stage}(b) represents a  pair $(\mathbf{X}_{1l},\mathbf{A}'_{l+1})$. For each such pair, choose one sequence $\mathbf{A}$ conditioned on the information at encoder $1$ at the end of block $(l+1)$. Similarly, for each right vertex $(\mathbf{X}_{2l},\mathbf{B}'_{l+1})$,  choose one sequence $\mathbf{B}$ at encoder $2$. The $\mathbf{A}$ and $\mathbf{B}$ sequences corresponding to $(\mathbf{X}_{1l},\mathbf{A}'_{l+1})$ and $(\mathbf{X}_{2l}, \mathbf{B}'_{l+1})$
are set to $\mathbf{A}_{l+2}$ and $\mathbf{B}_{l+2}$, respectively.

\item At the end of block $(l+2)$, the two encoders can decode $\mathbf{A}_{l+2}$ and $\mathbf{B}_{l+2}$ from $\mathbf{Y}_{l+2}$ with high probability. (The graph of Figure \ref{fig:four_stage}(b) should be sufficiently `thin' to ensure this). They now know the edge $(W_{1l},W_{2l})$, and the message graph is as shown in Figure \ref{fig:four_stage}(c). The two encoders cooperate to send $\mathbf{U}_{l+3}$ resolve the decoder's residual uncertainty.
\end{itemize}
Thus in this extended scheme, each message pair is decoded by the encoders with a delay of two blocks, and by the decoder with delay of one block.

\emph{Stationarity}:
  To obtain a single-letter rate region, we require a stationary   distribution of sequences in each block. In other words, we need the random sequences
  $(\mathbf{U},\mathbf{A}',  \mathbf{B}', \mathbf{A}, \mathbf{B}, \mathbf{X}_1, \mathbf{X}_2,\mathbf{Y})$ to be characterized by the same single-letter product distribution in each block. This will happen if we can ensure that the $\mathbf{A'},  \mathbf{B'}, \mathbf{A}, \mathbf{B}$ sequences in each block have the same single-letter distribution  $P_{A'B'AB}$.

  The correlation between $(\mathbf{A}'_{l+1},\mathbf{A}_{l+1})$ and $(\mathbf{B}'_{l+1},\mathbf{B}_{l+1})$ is  generated using the information available at each encoder at the end of block $l$. At this time,  both encoders know $\mathbf{s}_l \triangleq (\mathbf{u,a,b,y})_l$. In addition, encoder  $1$
  also knows $( \mathbf{a}'_l, \mathbf{x}_{1l})$  and hence we make it generate $(\mathbf{A', A})_{l+1}$ according to the product distribution
  $Q_{A'A|\tilde S \tilde{A}' \tilde X_1}^n(.|\mathbf{s}_{l}, \mathbf{a}'_l, \mathbf{x}_{1l} )$.   Recall that we use $\: \tilde{}\:$ to denote the sequence of the previous block.   Similarly, we make encoder $2$ generate  generate $(\mathbf{B', B})_{l+1}$ according to the product distribution
  $Q_{B'B|\tilde S \tilde{B}' \tilde X_2}^n(.|\mathbf{s}_{l}, \mathbf{b}'_l, \mathbf{x}_{2l})$.

  If the pair  $(Q_{A'A|\tilde S \tilde A' \tilde X_1 },Q_{B'B|\tilde S \tilde B' \tilde X_2})$ satisfy the   consistency condition defined below, the pair $(\mathbf{A',B', A,B})_{l+1}$  belongs to the typical set  $T(P_{A'B'AB})$  with high probability. This ensures stationarity of
  the coding scheme.  We state the coding theorem below.

\begin{defi} \label{def:consistency_better}
For a given MAC
$(\mathcal{X}_1,\mathcal{X}_2,\mathcal{Y},P_{Y|X_1,X_2})$ define
  $\mathcal{P}$ as the set of all distributions  $P$ on
$\mathcal{U} \times \mathcal{A} \times \mathcal{B}  \times \mathcal{A}' \times \mathcal{B}' \times
\mathcal{X}_1 \times \mathcal{X}_2 \times \mathcal{Y}$ of the form
\be
P_U P_{A'B'AB}P_{X_1|UA' A}P_{X_2|UB' B}P_{Y|X_1X_2}
\label{eq:joint_dist1_better}
\ee
where $\mathcal{U},\mathcal{A}',\mathcal{A} ,\mathcal{B}', \mathcal{B}$ are arbitrary finite
sets. Consider two sets of random variables $(U,A',B', A, B,X_1,X_2,Y)$ and
$(\tilde{U},\tilde{A}',  \tilde{B}', \tilde A, \tilde B, \tilde{X}_1,\tilde{X}_2,\tilde{Y})$
each having the above distribution $P$.
For conciseness, we refer to the collection $(U,A,B,Y)$ as $S$, and to
$(\tilde{U},\tilde A, \tilde B,\tilde{Y})$ as $\tilde{S}$. Hence
\[
P_{S,X_1,X_2}=P_{\tilde{S},\tilde{X}_1,\tilde{X}_2}=P.
\]
Define $\mathcal{Q}$ as the
set of  pairs of  conditional distributions
$(Q_{A'A|\tilde{S}, \tilde{A}' , \tilde{X}_1},Q_{B'B|\tilde{S} , \tilde{B}' , \tilde{X}_2})$ of the form
\begin{gather*}
Q_{A'A|\tilde{S}, \tilde{A}' , \tilde{X}_1} = Q_{A|\tilde{S}, \tilde{A}'} \cdot Q_{A'|A, \tilde{X}_1, \tilde{S}, \tilde{A}'} \\
Q_{B'B|\tilde{S}, \tilde{B}' , \tilde{X}_2} = Q_{B|\tilde{S}, \tilde{B}'} \cdot Q_{B'|B, \tilde{X}_2, \tilde{S}, \tilde{B}'}
\end{gather*}
that satisfy  the following consistency condition $\forall (a',b',a,b) \in \mathcal{A}' \times \mathcal{B}' \times \mathcal{A} \times \mathcal{B}$.
\be \label{eq:q1q2cond3stage}
\begin{split}
& P_{A'B'AB}(a',b',a,b) =  \hspace{-8pt} \sum_{\tilde{s},\tilde{a}', \tilde{b}', \tilde{x}_1,\tilde{x}_2}
\hspace{-10pt} \big[ P_{\tilde{S},\tilde{A}', \tilde{B}', \tilde{X}_1,\tilde{X}_2}(\tilde{s},\tilde{a}', \tilde{b}', \tilde{x}_1,\tilde{x}_2) \\
&\: \cdot Q_{A'A|\tilde{S},\tilde{A}',\tilde{X}_1}(a'\ a|\tilde{s},\tilde{a}' ,\tilde{x}_1)
Q_{B'B|\tilde{S}, \tilde{B}', \tilde{X}_2}(b'\ b|\tilde{s}, \tilde{b}' ,\tilde{x}_2) \big].
\end{split}
\ee
Then,  for any  $(Q_{A'A|\tilde{S},\tilde{A}',\tilde{X}_1},Q_{B'B|\tilde{S},\tilde{B}', \tilde{X}_2}) \in \mathcal{Q}$,
the joint distribution of the two sets of random variables - $(\tilde{S}, \tilde{A}', \tilde{B}', \tilde{X}_1,\tilde{X}_2)$ and
$({S},A', B', {X}_1,{X}_2)$ -  is
given by
\[
P_{\tilde{S} \tilde{A}' \tilde{B}' \tilde{X}_1 \tilde{X}_2}
Q_{A'A|\tilde{S},\tilde{A}',\tilde{X}_1}
Q_{B'B|\tilde{S},\tilde{B}', \tilde{X}_2}
P_{UX_1X_2Y|A'B' AB}.
\]
\end{defi}

\begin{thm} \label{thm:macfb_even_better}
For a MAC $(\mathcal{X}_1,\mathcal{X}_2,\mathcal{Y},P_{Y|X_1,X_2})$,
for any distribution $P$ from $\mathcal{P}$ and a pair of
conditional distributions $(Q_{A'A|\tilde{S},\tilde{A}',\tilde{X}_1},Q_{B'B|\tilde{S},\tilde{B}', \tilde{X}_2})$ from $\mathcal{Q}$,
the following rate-region is achievable.
\ben \label{eq:three_thm_statement}
\begin{split}
&R_1 <   I(X_1;Y|X_2B'B \tilde{S}U), \\
&R_1 < I(X_1;Y|X_2A'B'A B \tilde S U)+I(A';Y|B'A B \tilde S U) \\
& \qquad +I(A;Y|B \tilde S U)+I(U;Y), \\
&R_2 <  I(X_2;Y|X_1A'A \tilde{S}U),\\
&R_2 <  I(X_2;Y|X_1 A'B' A B \tilde{S} U)+I(B';Y|A'A B \tilde{S} U)\\
& \qquad +I(B;Y|A \tilde S U)+I(U;Y), \\
& R_1+R_2 < I(X_1X_2;Y|U \tilde S)+I(U;Y).
\end{split}
\een
\end{thm}

The proof essentially consists of: a) Computing the left and right degrees of the message graph  at each stage in Figure \ref{fig:four_stage},
b) ensuring both encoders can decode $(\mathbf{A},\mathbf{B})$ (the edge from the graph \ref{fig:four_stage}(b)) in each block, and
c) ensuring that the decoder can decode $\mathbf{U}$ in each block.

We omit the formal proof since it is an extended version of the arguments in  Section \ref{sec:proof}.

\section{Conclusion}
\label{sec:conclusion}
We proposed a new single-letter achievable rate region for the two-user discrete memoryless MAC with noiseless feedback. This rate region is achieved through a block-Markov superposition coding scheme, based on the observation that the messages of the two users are correlated given the feedback. We can represent the messages of the two users as left and right vertices of a bipartite graph. Before transmission, the graph is fully connected, i.e., the messages are independent. The idea is to use the feedback  to thin the graph gradually, until it reduces to a set of disjoint edges.  At this point, each encoder knows the message of the other, and they  can cooperate to resolve the decoder's residual uncertainty. It is not clear if this idea can be applied to a MAC  with partial/noisy feedback - the difficulty lies in identifying common information between the encoders to summarize at the end of each block.  However, this method of exploiting correlated information could be useful in other multi-terminal communication problems.

\appendix
\section*{Computing the symmetric sum rate} \label{app:compute}
\begin{figure*}[!t]
\normalsize
\begin{equation*}
\begin{split}
& H(Y|X_2ABU) =  x \sum_{u} p_u [(1-p_{0u}) h(q p_{1u}) +p_{0u} h(q(1+p_{1u})) + (1-p_{1u}) h(q p_{0u}) + p_{1u} h(q(1+p_{0u}))]\\
& \qquad + y \sum_{u}p_u [(1-p_{1u}) h(q p_{1u}) + p_{1u} h(q(1+p_{1u}))] + (1-2x-y) \sum_{p_u}[ (1-p_{0u}) h(q p_{0u}) + p_{0u} h(q(1+p_{0u})) ], \\
\end{split}
\end{equation*}
{\small{
\begin{equation*}
H(Y|UB \tilde Y \tilde X_2) = \sum_{u} p_u \left[ (x+y) h\left( q \: (p_{u1} +  \frac{x p_{u0} + y p_{u1}}{x+y})\right)
 +  (1-x-y) h\left( q \: (p_{u0}+ \frac{(1-2x-y) p_{u0} +  x p_{u1}}{1-x-y}) \right)\right] + o(q),
\end{equation*}
}}
\begin{equation*}
\begin{split}
& H(Y|UB X_2 \tilde Y \tilde X_2) = \sum_u p_u (x+y)\left( p_{u1} h\left( q \: (1+ \frac{x p_{u0} +  y p_{u1}}{x+y})\right)
+ (1-p_{u1}) h\left( q \: \frac{x p_{u0} + y p_{u1}}{x+y}\right) \right)\\
& \qquad + \sum_u p_u (1-x-y)\left( p_{u0} h\left( q \: ( 1+ \frac{(1-2x-y) p_{u0} +  x p_{u1}}{1-x-y})\right)
+ (1-p_{u0}) h\left( q \:\frac{(1-2x-y) p_{u0} +  x p_{u1}}{1-x-y}\right) \right) + o(q).
\end{split}
\end{equation*}
\hrulefill
\vspace*{2pt}
\end{figure*}
The random variables $U,A,B,X_1,X_2$ are all chosen to have binary alphabet. The stationary input distribution has the form $P_U \cdot P_{AB} \cdot P_{X_1|AU} \cdot P_{X_2|BU}$ and is defined as follows.
\begin{align}
&P_U(0)=p_0, \quad P_U(1)=p_1=1-p_0, \\
& P_{AB}(1,1)=y, \quad P_{AB}(0,1)=P_{AB}(0,1)=x, \nonumber \\
& P_{AB}(0,0)=1-2x-y, \label{eq:ab}\\
&P_{X_1|UA}(1|u,0)=P_{X_1|UB}(1|u,0)=p_{u0}, \nonumber \\
& P_{X_1|UA}(1|u,1) = P_{X_2|UB}(1|u,1)=p_{u1}, \quad u \in \{0,1\}.
\end{align}

Recall that $\tilde S = (\tilde U, \tilde A, \tilde B, \tilde Y)$. The distributions $Q_{A|\tilde X_1 \tilde S}$ and $Q_{B|\tilde X_1 \tilde S}$, which generate $A$ and $B$ using the feedback information, are defined as follows.
\be \label{eq:gen_A}
Q_{A|\tilde X_1 \tilde S}:  \quad A = \left\{
\begin{array}{cc}
1 \text{ if } \tilde{X}_1 \neq \tilde Y \\
0 \text{ if } \tilde{X}_1 = \tilde Y
\end{array}
\right.
\ee
\be \label{eq:gen_B}
Q_{B|\tilde X_2 \tilde S}: \quad  B = \left\{
\begin{array}{cc}
1 \text{ if } & \tilde{X}_2 \neq \tilde{Y} \\
0 \text{ if } & \tilde{X}_2 = \tilde Y
\end{array}
\right.
\ee

For \eqref{eq:gen_A} and \eqref{eq:gen_B} to generate a joint distribution $P_{AB}$ as in \eqref{eq:ab}, the consistency condition given by \eqref{eq:q1q2cond} needs to be satisfied.
Thus we need
\begin{align}
&P_{AB}(1,1) =  y = P(\tilde X_1 = 1, \tilde X_2 =1, \tilde Y =0) \label{eq:y_c} \\
&P_{AB}(0,1) =  x \nonumber \\
&= P(\tilde X_1 = 0, \tilde X_2 =1, \tilde Y =0) + P(\tilde X_1 = 1, \tilde X_2 =0, \tilde Y =1) \label{eq:xc_1} \\
&P_{AB}(1,0) =  x \nonumber \\
&= P(\tilde X_1 = 1, \tilde X_2 =0, \tilde Y =1) + P(\tilde X_1 = 0, \tilde X_2 =1, \tilde Y =0). \label{eq:xc_2}
\end{align}
We can expand \eqref{eq:y_c} as
\begin{equation}
\begin{split}
&y = P(\tilde X_1 = 1, \tilde X_2 =1) (1-q)\\
&= \sum_{u} p_u \, ( y  p_{u1}^2 + 2x p_{u0}p_{u1} + (1-2x-y) p^2_{u0} )\,(1-q).
\end{split}
\end{equation}
As $q \to 0$, the above condition becomes
\be \label{eq:y_cond}
y= \sum_u p_u ( y p_{u1}^2 + 2x  p_{u0}p_{u1} + (1-2x-y) p^2_{u0} ).
\ee
Similarly, as $q \to 0$, \eqref{eq:xc_1} and \eqref{eq:xc_2} become
\begin{equation} \label{eq:x_cond}
\begin{split}
x= \sum_u & p_u [y (1-p_{u1}) p_{u1} + x  (1-p_{u1}) p_{u0} \\
&+ x (1-p_{u0}) p_{u1} + (1-2x-y)(1-p_{u0})p_{u0} ].
\end{split}
\end{equation}
\eqref{eq:x_cond} and \eqref{eq:y_cond} can be written in matrix form as
\be \label{eq:matrix_form}
\begin{bmatrix} a_{11} & a_{12} \\ a_{21} & a_{22} \end{bmatrix}
\begin{bmatrix} x \\ y \end{bmatrix}=
\begin{bmatrix} \sum_u p_u  p_{u0}(1-p_{u0}) \\ \sum_u p_u \ p_{u0}^2\end{bmatrix}
\ee
where
\begin{equation*}
\begin{split}
&a_{11} \triangleq 1 - \sum_u p_u (p_{u1}- p_{u0}) (1-2p_{u0}), \\
&a_{12} \triangleq \sum_u p_u [ p_{u0}(1-p_{u0}) -  p_{u1}(1-p_{u1})], \\
&a_{21} \triangleq 2 \sum_{u} p_u p_{u0} (p_{u0}-p_{u1}), \; a_{22} \triangleq 1- \sum_{u} p_u (p_{u1}^2 - p_{u0}^2).
\end{split}
\end{equation*}
\eqref{eq:matrix_form} uniquely determines $x$ and $y$ given the values of  $p_u, p_{u0}$ and $p_{u1}$   for $u \in \{0,1\}$.
Therefore the joint distribution is completely determined.
\subsection*{The information quantities}
We calculate the information quantities in nats below. We use the notation $h(.)$ to denote the binary entropy function in nats.
\be
h(x) = -x \ln x - (1-x) \ln (1-x), \quad 0\leq x \leq 1.
\ee
\ben
\begin{split}
& H(Y)  = h(2q(x+y)),\\
& H(Y|U)  =  \sum_{u=0}^1 p_u \cdot h(2q((x+y)p_{u1}+(1-x-y)p_{u0})),
\end{split}
\een
\ben
\begin{split}
& H(Y|X_1X_2)  =  2x h(q) + y h(2q), \\
& H(Y|ABU) =  \sum_{u=0}^1 p_u [ 2x h(q(p_{u1}+p_{u0})) + y h(2q p_{u1}) \\
& \hspace{1.2in} + (1-2x-y) h(2q p_{u0})], \\
& H(Y|\tilde Y U) = H(Y|U) + o(q),
\end{split}
\een
 and $H(Y|X_2ABU)$, $H(Y|UB\tilde{Y} \tilde{X}_2)$, $H(UBX_2 \tilde{Y} \tilde{X}_2)$ are given by the equations at the top of this page.
Here $o(q)$ is any function such that $\frac{o(q)}{q} \to 0$ as $q \to 0$. Using these in the rate constraints of \eqref{eq:achievable_region}, we can obtain the constraints for $R_1$ and $R_1+R_2$. Due to the symmetry of the input distribution, the bound for $R_2$ is the same as that for $R_1$ above.
Optimizing over $p_u,p_{u0},p_{u1}$ for $u \in \{0,1\}$, we obtain an achievable symmetric sum rate of
\ben
R_1+R_2 = 0.5132 q + o(q) \text{ nats }
\een
for
\begin{align*}
&P(U=0)=p_0=0.0024, \; P(U=1)=1-p_0=0.9976, \\
&P_{X_1|UA}(1|0,0)=p_{00}=0.791, \\
&P_{X_1|UA}(1|1,0)=p_{10}=\epsilon, \; (\epsilon \text{ is a constant very close to }0), \\
&P_{X_1|UA}(1|0,1)=p_{01}=0.861, \\
&P_{X_1|UA}(1|1,1)=p_{11}=0.996.
\end{align*}

\section*{Acknowledgements}
We thank the anonymous reviewers and the associate editor for their valuable comments, which led to a significantly improved paper.

\IEEEtriggeratref{20}

\vspace{-5in}

\begin{IEEEbiographynophoto}{Ramji Venkataramanan}
received the B.Tech degree in Electrical Engineering from the Indian Institute of Technology, Madras in 2002, and the Ph.D degree in Electrical Engineering (Systems) from the University of Michigan, Ann Arbor in  2008. He is currently a postdoctoral research associate at Yale University.
His research interests include information theory, coding and stochastic network theory.
\end{IEEEbiographynophoto}

\vspace{-5in}

\begin{IEEEbiographynophoto}{S. Sandeep Pradhan}
 obtained his M.E. degree from the Indian Institute of Science in 1996 and Ph.D. from the University of California at Berkeley in 2001. From 2002 to 2008 he was an assistant professor in the Department of Electrical Engineering and Computer Science at the University of Michigan at Ann Arbor, where he is currently an associate professor. He is the recipient of 2001 Eliahu Jury award given by the University of California at Berkeley for outstanding research in the areas of systems, signal processing, communications and control, the CAREER award given by the National Science Foundation (NSF), and the Outstanding achievement award for the year 2009 from the University of Michigan. His research interests include sensor networks, multi-terminal communication systems, coding theory, quantization, information theory.
\end{IEEEbiographynophoto}

\end{document}